%% file: paper.tex
\pdfoutput=1
\documentclass[12pt]{article}

\usepackage{jheppub}
\usepackage{mathrsfs}
\usepackage{psfrag}

\usepackage{epstopdf}

\usepackage[table]{xcolor}
\usepackage{amsmath,bbm,array,amsfonts,graphicx,wrapfig,lscape,float,slashbox,multirow,longtable,rotating,subfigure}
\usepackage{url}

\input{pref}
\newcolumntype{C}[1]{>{\centering\arraybackslash}m{#1}}

\newcommand{\bb}{\mathsf{b}}
\newcommand{\ww}{\mathsf{w}}

\newcommand{\Gr}{$Gr_{k,n}$}
\newcommand{\pl}{Pl\"ucker }

\title{Bipartite Field Theories, Cluster Algebras and the Grassmannian}
\author{Sebasti\'an Franco,}
\author{Daniele Galloni,}
\author{Alberto Mariotti}

\affiliation{
Institute for Particle Physics Phenomenology, Department of Physics\\
Durham University, Durham DH1 3LE, United Kingdom
}

\emailAdd{sebastian.franco@durham.ac.uk, daniele.galloni@durham.ac.uk,alberto.mariotti@durham.ac.uk}

\abstract{We review recent progress in Bipartite Field Theories. We cover topics such as their gauge dynamics, emergence of toric Calabi-Yau manifolds as master and moduli spaces, string theory embedding, relationships to on-shell diagrams, connections to cluster algebras and the Grassmannian, and applications to graph equivalence and stratification of the Grassmannian.}

\preprint{
\begin{flushright}IPPP/14/30\end{flushright} \vspace{-0.9cm}
\begin{flushright}DCPT/14/60\end{flushright}
}

\begin{document}

\maketitle

%===============================================================================
%===============================================================================

%===============================================================================
\section{Quantum Field Theory, Duality and Combinatorics}
%===============================================================================

Quantum field theory (QFT) underlies our description of fields as diverse as particle physics, statistical mechanics and condensed matter physics. It leads to extremely precise predictions which are constantly tested experimentally. It is thus remarkable that despite it being such a fundamental and mature framework our understanding of QFT is currently undergoing tremendous progress. This progress occurs along multiple fronts, including holography, integrability and duality, to name a few. 

Powerful mathematical and geometric ideas play a central role in some of the most recent developments. A common theme in recent years has been the definition of QFTs in terms of some underlying geometric or combinatorial objects. In these constructions, it is often possible to construct theories by assembling certain geometric elementary building blocks, which have gauge theory counterparts. Furthermore, gauge theory dualities are captured by basic transformations of the underlying geometric objects. 

The connection between dimer models and 4d $\mathcal{N}=1$ quiver gauge theories on systems of D-branes probing  toric Calabi-Yau 3-folds is one example that falls in this category \cite{Hanany:2005ve,Franco:2005rj,Franco:2005sm,Kennaway:2007tq}. Dimer models, i.e.\ bipartite graphs on a 2-torus, are indeed a subset of the general class of theories discussed in this review.  Among other things, they have been instrumental in the discovery of the first infinite families of explicit AdS/CFT dual pairs in 4d \cite{Franco:2005sm,Butti:2005sw}. In addition, they provide the largest known classification of purely $\mathcal{N}=1$ SCFTs in 4d. 

\begin{figure}[h]
\begin{center}
\includegraphics[width=11cm]{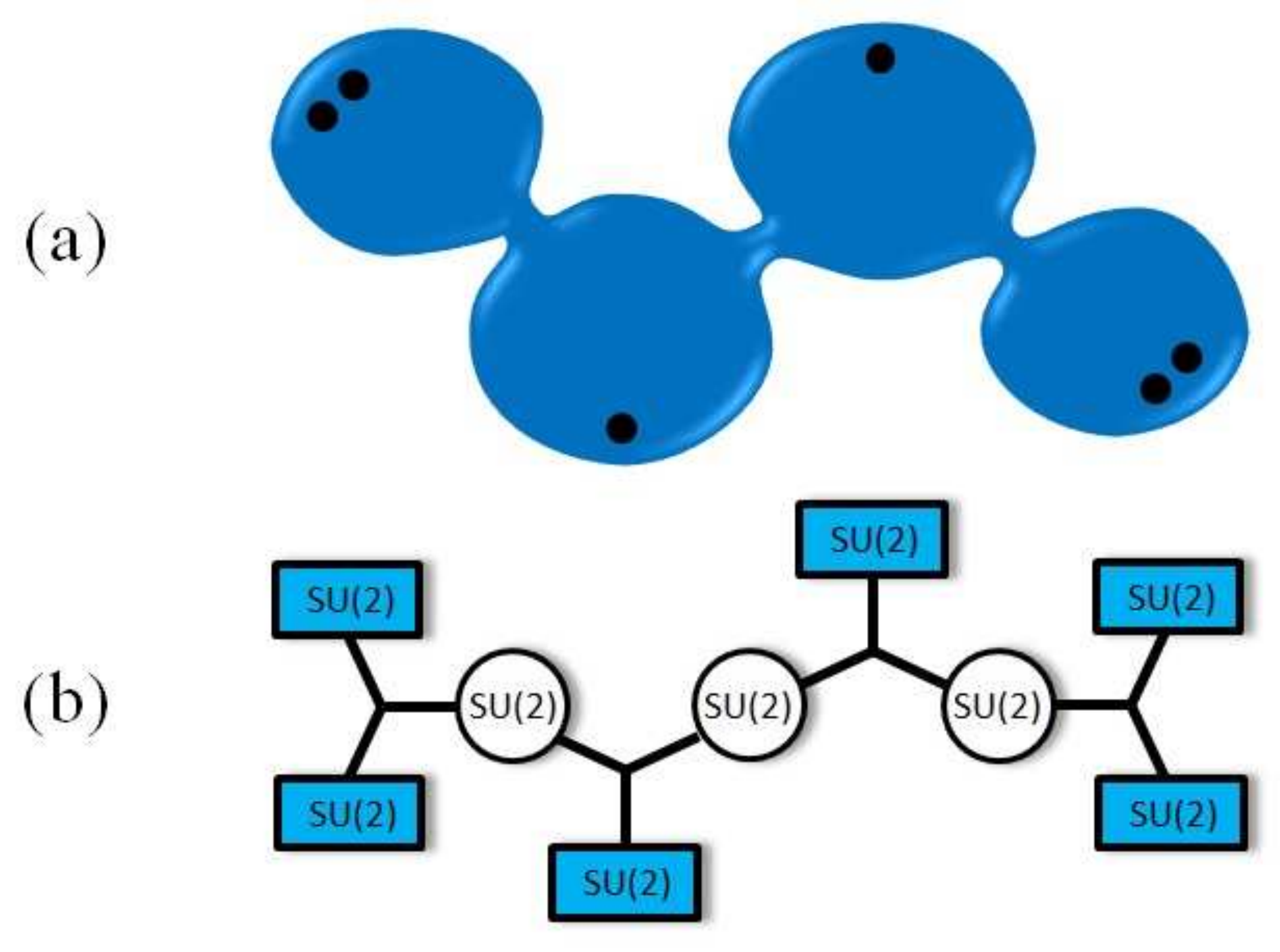}
\caption{a)	A pictorial representation of a punctured Riemann surface giving rise to a 4d $\mathcal{N}=2$ Gaiotto theory. b) The corresponding generalized quiver obtained from compactifying the $A_1$ $(2,0)$ 6d SCFT on it. Circles indicate gauge symmetries while rectangles correspond to global symmetries.} 
\label{Gaiotto}
\end{center}
\end{figure}

Another paradigmatic example of this general approach is provided by Gaiotto dualities in 4d $\mathcal{N}=2$ gauge theories \cite{Gaiotto:2009we}. These theories are constructed by compactifying a $(2,0)$ SCFT of $A_{N-1}$ type
on a punctured Riemann surface. In this construction, different pants decompositions of the Riemann surface correspond to different S-duality frames of the gauge theories. Remarkably, it is possible to make precise statements about dualities, despite the fact that these theories do not even have a Lagrangian description and the matter building blocks are strongly coupled. \fref{Gaiotto} shows an example of a punctured Riemann surface giving rise to one such theory and the generalized quiver associated to the specific pants decomposition. This general approach has been fruitfully extended to theories in various dimensions and with different amounts of SUSY. An incomplete list of examples includes \cite{Benini:2009mz,Bah:2011je,Bah:2012dg,Benini:2010uu,Benini:2013cda}.

%===============================================================================
%===============================================================================

A final example of the geometric approach to QFTs, involves the construction of 3d gauge theories in terms of 3-dimensional manifolds \cite{Dimofte:2011ju}. In this class of models, different triangulations of the 3-manifold map to mirror symmetry of the gauge theories.

This article is devoted to Bipartite Field Theories (BFTs), a new class of quantum field theories with a combinatorial construction similar to the examples presented above. It is intended as an introductory review of the recent progress in BFTs, aimed at both a physicist and mathematician audience. It will hopefully serve as a quick reference for the main concepts and spark interest in possible directions for future research. We encourage the reader to visit the references for detailed presentations of the topics.

\bigskip

%===============================================================================
\section{Bipartite Field Theories}
%===============================================================================

\label{section_BFTs}

Bipartite field theories are 4d $\mathcal{N}=1$ supersymmetric gauge theories whose Lagrangians are defined by bipartite graphs
embedded into a Riemann surface, possibly with boundaries \cite{Franco:2012mm}.\footnote{Closely related theories were introduced in \cite{Xie:2012mr}.}

4d $\mathcal{N}=1$ supersymmetric gauge theories are determined by 
specifying the gauge symmetry group (vector superfields), matter content (chiral superfields), a real function of the chiral superfields 
(the K\"{a}hler potential) and a holomorphic function of the chiral superfields (the superpotential $W$). In this article we will focus on theories with canonical K\"{a}hler potential, so we will not mention it any longer.

A bipartite graph is a graph in which nodes can be colored white or black,
such that white nodes are connected only to black nodes and vice-versa. Once a bipartite graph is embedded into a Riemann surface, 
nodes can be further distinguished into internal and external. External nodes are those sitting on the boundaries of the Riemann surface. A further characterization of a node is its valence, the number of edges connected to it. In our construction, we allow only external nodes with valence one.

Faces are regions on the Riemann surface which are
surrounded by edges and/or boundaries. They can also be divided into two disjoint categories, external or internal. External faces are the ones whose 
perimeter includes at least one boundary. 

The map defining a BFT in terms of a bipartite graph on a Riemann surface is succinctly explained in Table \ref{tab:dictionaryBFT}.\footnote{It is possible and indeed well motivated to consider the case in which the ranks of the $U(N_i)$ symmetry groups for faces are not all equal. For simplicity, we will not contemplate this possibility in this article.}

%===============================================================================
\begin{table}[h]
\begin{center}
\begin{tabular}{|p{.25\textwidth}|p{.50\textwidth}|}
\hline
{\bf Graph} & {\bf BFT} \\
\hline
\hline
Internal Face ($2n$ sides) & $U(N)$ gauge symmetry group with $n \times N$ flavors. \\
\hline
External Face & $U(N)$ global symmetry group \\
\hline
Edge between faces $i$ and $j$ & Chiral superfield in the bifundamental representation of 
groups $i$ and $j$ (adjoint representation if $i=j$). The chirality, i.e.\ orientation, of the bifundamental is such that  it goes clockwise around white nodes and counter-clockwise around black nodes.
\\
\hline
$k$-valent node & Superpotential term made of $k$ chiral superfields. Its sign is $+/-$ for a white/black node, respectively. \\
\hline
\end{tabular}
\end{center}
\caption{The dictionary relating bipartite graphs on Riemann surfaces to BFTs.}
\label{tab:dictionaryBFT}
\end{table}
%===============================================================================

\begin{figure}[h]
\begin{center}
\includegraphics[width=12cm]{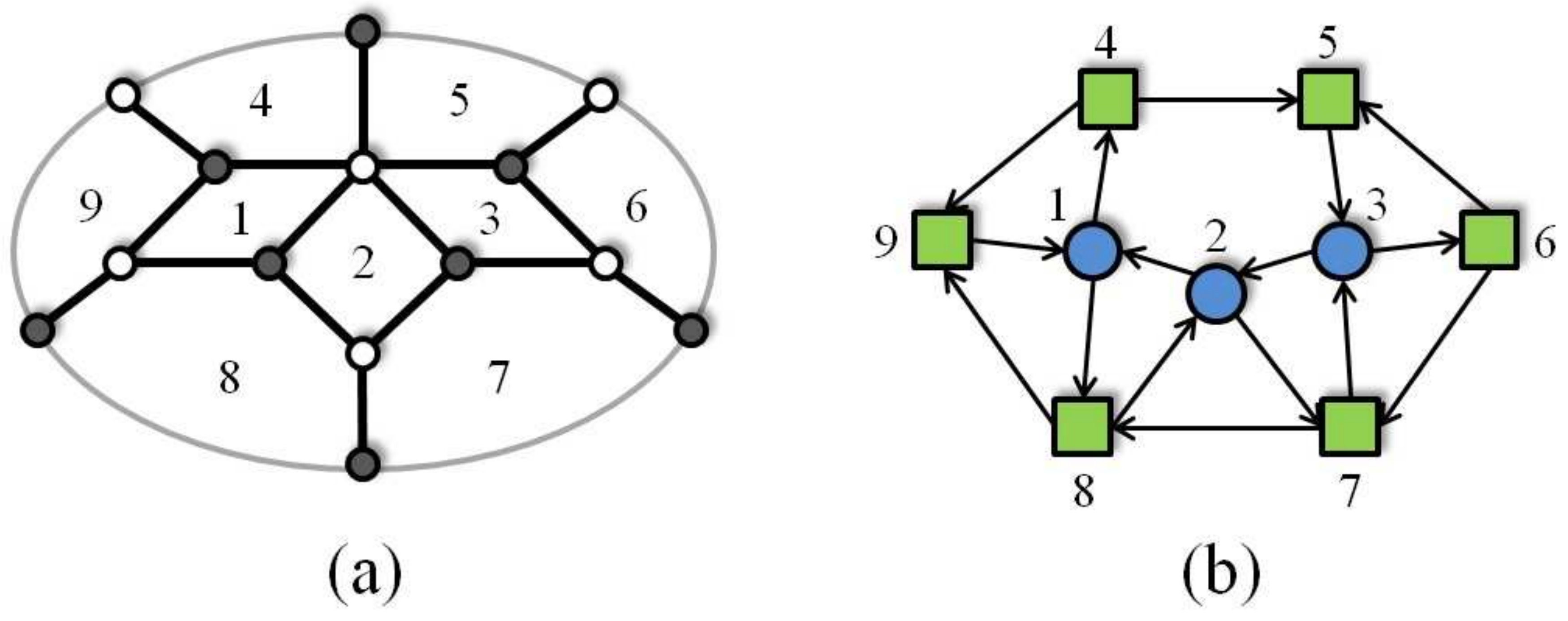}
\caption{A bipartite graph on a disk and its dual quiver. Every plaquette in this quiver corresponds to a node in the original graph and hence a superpotential term.}
\label{dual_quiver}
\end{center}
\end{figure}

We can alternatively think about these theories in terms of a quiver dual to the graph, as illustrated in \fref{dual_quiver}. This quiver is such that its plaquettes, i.e.\ the minimal oriented closed loops, encode the terms in the superpotential of the BFT.

%===============================================================================
%===============================================================================
 
A special subclass  of BFTs, the one defined on a torus without boundaries, also known as {\it brane tilings}, describes the worldvolume theory of D3-branes probing toric Calabi-Yau singularities in type IIB String Theory \cite{Hanany:2005ve,Franco:2005rj,Franco:2005sm,Kennaway:2007tq}. They constitute an infinite class of $\mathcal{N}=1$ SCFT which have been a remarkable testing ground for the AdS/CFT duality \cite{Benvenuti:2004dy,Franco:2005sm,Butti:2005sw,Benvenuti:2005ja}.

For non-planar graphs, an alternative possibility for gauging symmetries was introduced in \cite{Franco:2012wv}. The theories resulting from this choice should be regarded as a new class of BFTs. Interestingly, this new class of theories becomes independent of an underlying Riemann surface. While extremely interesting and relevant for the non-planar examples we mention, we will not use this gauging for any explicit computation in this review.

We refer those interested in extensive catalogues of explicit BFT examples to \cite{Franco:2012mm,Franco:2012wv} for general BFTs and to \cite{Hanany:2012vc,Cremonesi:2013aba,He:2014jva} for higher genus examples without boundaries.

\bigskip

%===============================================================================
\subsection{Graphical Gauge Theory Dynamics}
%===============================================================================

\label{section_graphical_dynamics}

\begin{figure}[h]
\begin{center}
\includegraphics[width=14cm]{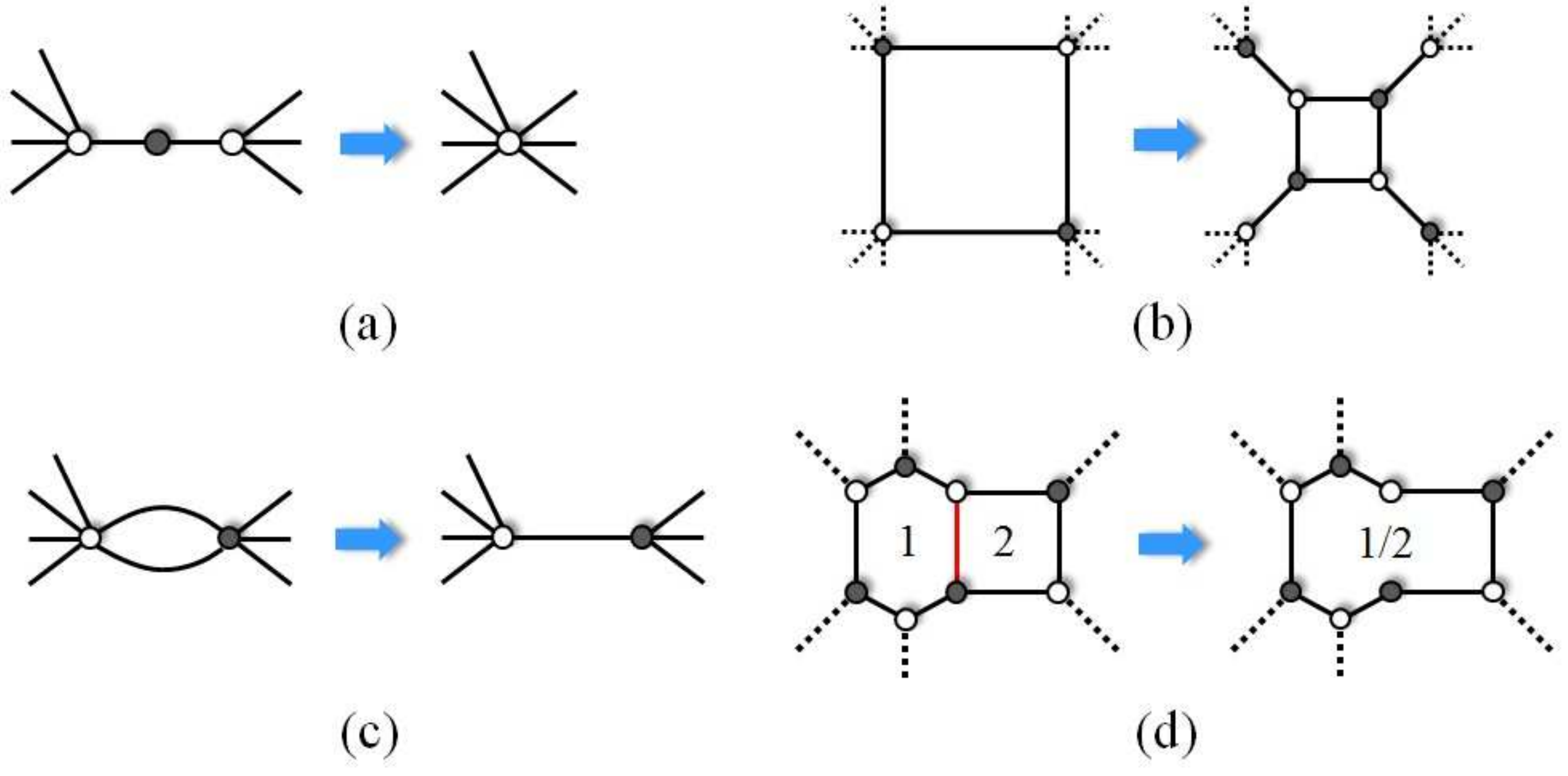}
\caption{Graph transformation encoding several dynamical processes in the gauge theory.}
\label{graphical_dynamics}
\end{center}
\end{figure}

Since BFTs are fully determined in terms of a graph on a Riemann surface, their dynamics translates into transformations of the graph. \fref{graphical_dynamics} shows the operations associated to the following gauge theory processes:

\medskip

\begin{itemize}
\item[{\bf (a)}] Integrating out massive fields.
\item[{\bf (b)}] Seiberg duality \cite{1995NuPhB.435..129S} of an $N_f=2N_c$ gauge group.
\item[{\bf (c)}] Confinement of an $N_f=N_c$ gauge group and going to the branch of the quantum moduli space on which mesons do not get a non-vanishing vacuum expectation value.
\item[{\bf (d)}] Higgs mechanism.\footnote{In general, the Higgs mechanism corresponds to removing an edge between two adjacent faces. \fref{graphical_dynamics} shows one explicit example of this situation. If the removed edge sits between two external faces, i. the process just corresponds to spontaneous symmetry breaking of global symmetries.}
\end{itemize}

\medskip

For a comprehensive discussion of these processes we refer the reader to \cite{Franco:2012mm,Xie:2012mr}.

%===============================================================================
%===============================================================================
\bigskip

%===============================================================================
\section{Bipartite Technology}
%===============================================================================

\label{sec:BipaTec}

In this section we quickly review various types of objects which are useful in the study of bipartite graphs.

\bigskip
 
%=============================================================================== 
 \paragraph{Perfect Matchings.}
%===============================================================================
 
These objects play a fundamental role in the analysis of bipartite graphs. A perfect matching is a collection of the edges in a bipartite graph such that each internal node is the endpoint of exactly one edge in the perfect matching and external nodes are the endpoints of one or zero edges in it.\footnote{Strictly speaking, the objects we have just defined are called {\it almost perfect matchings}. For brevity, we will refer to them as perfect matchings throughout the article.}

\begin{figure}[h]
\begin{center}
\includegraphics[width=14cm]{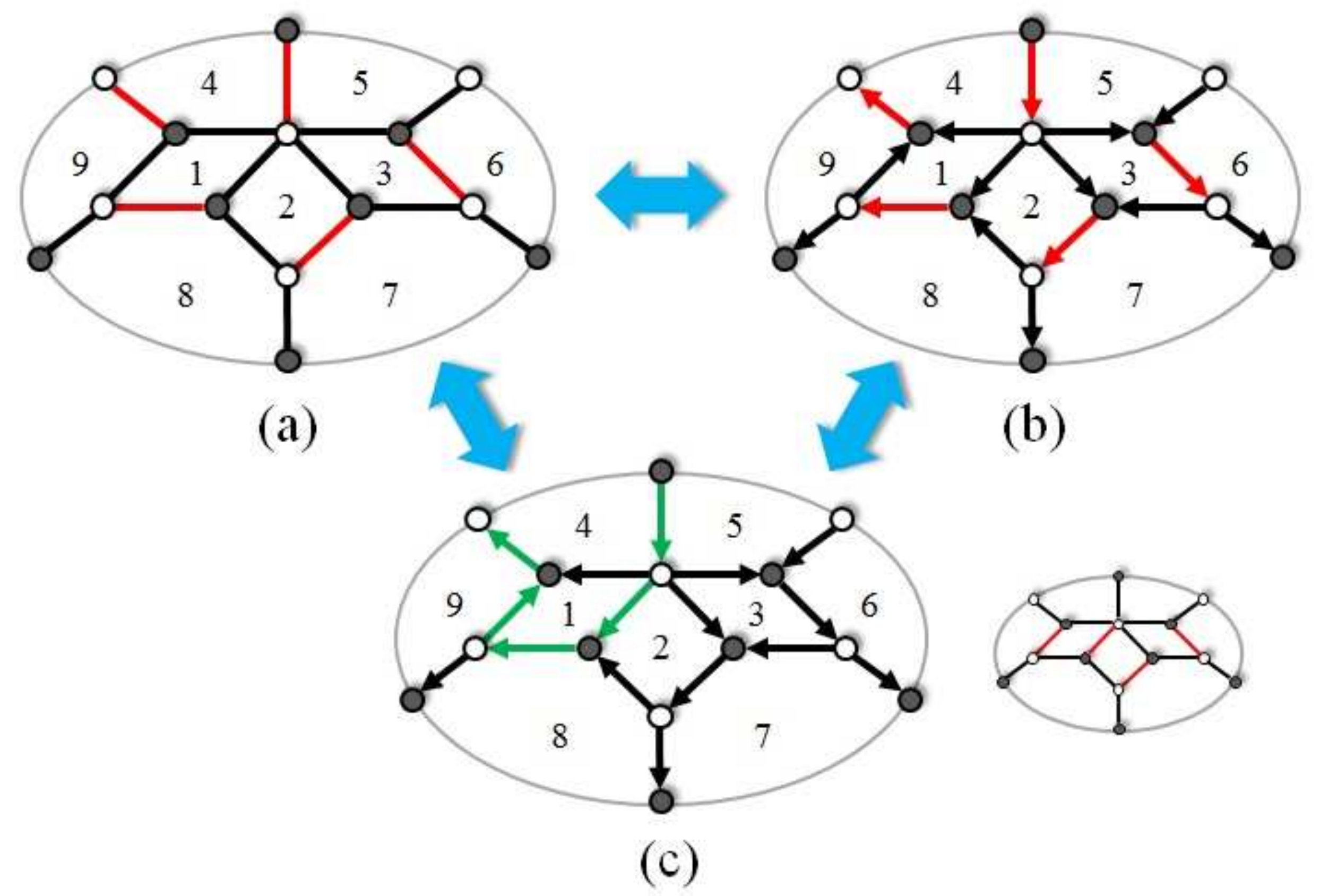}
\caption{An example of: a) a perfect matching, b) the corresponding perfect orientation, and c) a flow in this perfect orientation, corresponding to the perfect matching shown on the right.}
\label{pms_orientations_flows}
\end{center}
\end{figure}

Generally, there are multiple perfect matchings for a given bipartite graph.  There is a very efficient procedure for determining all perfect matching of a bipartite graph, even including boundaries, based on a generalization of the Kasteleyn matrix (see \cite{Franco:2012mm} for a detailed explanation). \fref{pms_orientations_flows}.a  shows an example of a perfect matching for the graph in \fref{dual_quiver}.a. In fact, this graph gives rise to $25$ different perfect matchings $p_\mu$, which we list in Appendix \ref{app:PMlist}.

As we will explain in \sref{section_master_and_moduli_spaces}, perfect matchings provide ideal variables for describing the moduli space geometry of the corresponding BFT.

\bigskip
 
%===============================================================================
\paragraph{Perfect Orientations.}
%===============================================================================
A perfect orientation is an assignment of arrows for every edge of the graph such that white $k$-valent nodes have exactly 1 incoming and $k-1$ outgoing arrows, while black $k$-valent nodes have 1 outgoing and $k-1$ incoming arrows. \fref{pms_orientations_flows}.b shows  a perfect orientation. There is a one-to-one map between perfect matchings and perfect orientations; given a perfect matching, its edges indicate the single incoming/outgoing arrows at white/black nodes.

In the presence of boundaries, every perfect orientation divides external nodes into sources and sinks. Note, however, that different perfect orientations on a given graph can give rise to the same sets of sources and sinks.

\bigskip
 
%===============================================================================
\paragraph{Flows.} 
%===============================================================================
Given a perfect orientation, we refer to the oriented paths in it as {\it flows} and we denote them $\mathfrak{p}_{\mu}$. Flows might consist of more than one disjoint component. Once again, it is possible to establish a bijection between flows and perfect matchings. Following the previous discussion, 
the underlying perfect orientation corresponds to a perfect matching, which we call the reference perfect matching $p_{\text{ref}}$. Then, every flow in the perfect orientation can be obtained by subtracting the reference perfect matching from another perfect matching. When doing so, the edges in $p_{\text{ref}}$ are taken with black to white orientation, while the ones belonging to the other perfect matching are given a white to black orientation. Edges which are common to both perfect matchings cancel and disappear from the flow. The reference perfect matching is mapped by this procedure into the trivial flow, i.e.\ a flow that does not involve any edge.

\smallskip
\bigskip

\fref{pms_orientations_flows} illustrates the connection between the perfect matchings, perfect orientations and flows.

%===============================================================================
%===============================================================================

\bigskip

%===============================================================================
\section{Three Routes into Polytopes and Toric Geometry} \label{3polytopeRoutes}
%===============================================================================

\begin{figure}[h]
\begin{center}
\includegraphics[width=13cm]{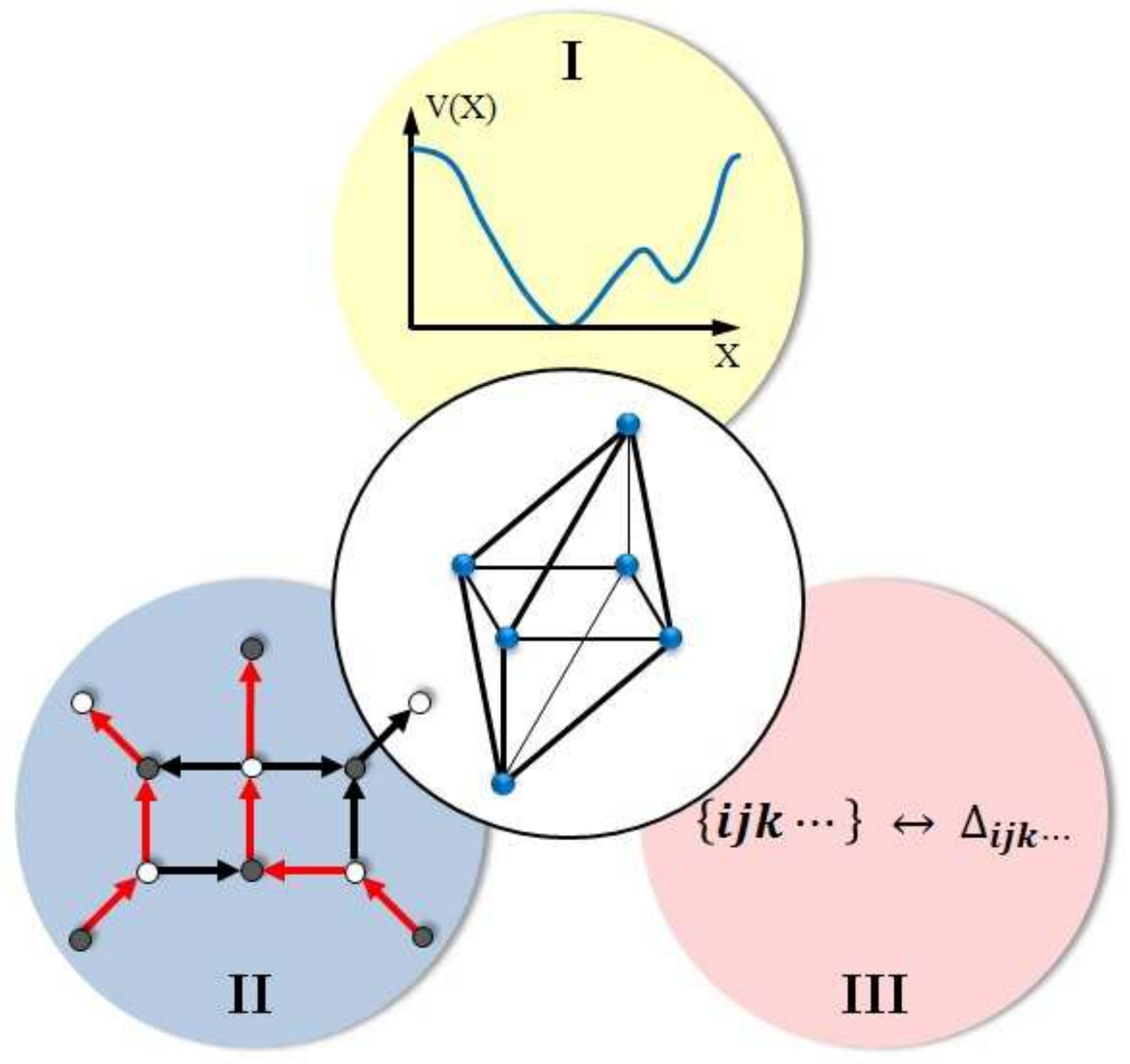}
\caption{Three alternative approaches for obtaining the polytopes of interests. I) As the toric diagrams of the master and moduli spaces of BFTs. II) From a geometric descriptions of flows in a perfect orientation. III) From matroids.}
\label{3_to_polytope}
\end{center}
\end{figure}

Every bipartite graph is associated to a pair of polytopes, which can also be interpreted as toric diagrams defining certain non-compact, singular, toric Calabi-Yau (CY) manifolds. For planar graphs, they are called {\it matching} and {\it matroid polytopes}. For simplicity, we will often call their non-planar counterparts by the same names. In this section we review three superficially different, but in fact equivalent, ways of arriving at such polytopes, as summarized in \fref{3_to_polytope}.\footnote{The examples in this figure have been chosen for purely pictorial reasons. They do not correspond to each other.} As usual, having multiple perspectives on a given object often leads to powerful insights. The different methods will be illustrated using the example in \fref{dual_quiver}.a.

This section is based on \cite{Franco:2012mm, Franco:2013nwa}. Related material, sometimes restricted to planar graphs, has also appeared in \cite{2007arXiv0706.2501P,2008arXiv0801.4822T,Amariti:2013ija}.

%===============================================================================
%===============================================================================

\bigskip

%===============================================================================
\subsection{Polytopes from Master and Moduli Spaces}
%===============================================================================

\label{section_master_and_moduli_spaces}

The {\it moduli space} is a fundamental object in the study of SUSY QFTs. It is the space of field configurations for which the scalar potential of the theory vanishes. Since the scalar potential is positive semi-definite in SUSY theories, the points where it vanishes correspond to absolute minima. Moreover, in SUSY theories such vacua are not isolated, and the moduli space becomes an interesting continuous geometry.

The scalar potential is a sum of two types of contributions: F-terms (associated to chiral superfields) and D-terms (associated to vector superfields). Every F- and D-term enters the scalar potential with squared absolute value and hence needs to independently vanish on the moduli space. We will not discuss the basics of F- and D-terms in this review. 

In the rest of the paper, we will focus on classical, Abelian BFTs. A remarkable feature of BFTs is that due to the very special structure of their superpotential, which follows from a bipartite graph according to the dictionary given in \sref{section_BFTs}, the determination of the moduli space reduces to a combinatorial problem of perfect matchings.

The moduli space can be constructed in two stages. First, we construct the space of vanishing F-terms, which is known as the {\it master space} of the theory \cite{Forcella:2008bb}. In doing so, we only require vanishing of the F-terms for fields associated to internal edges in the bipartite graph, without imposing zero F-terms for external edges. There are both geometric and physical reasons motivating this treatment, as explained in detail in \cite{Franco:2012mm}. We will reflect on the latter in \sref{section_string_embedding}.

Every internal edge $X_0$ appears in exactly two terms in the superpotential, which thus takes the general form
\beq
W=X_0 P_1(X_i)- X_0 P_2(X_i) +\ldots,
\eeq
where $P_1(X_i)$ and  $P_2(X_i)$ represent products of bifundamentals fields. The F-term equation for every $X_0$ is given by

\beq
\partial_{X_0}W =0 \ \ \ \iff \ \ \ P_1(X_i)=P_2(X_i).
\label{F_internal}
\eeq

As shown in \cite{Franco:2012mm}, all F-term equations \eref{F_internal} are automatically solved by the following change of variables
\beq
X_i=\prod_\mu p_\mu^{P_{i\mu}},
\label{X_from_p}
\eeq
where
\beq
P_{i\mu}=\left\{ \begin{array}{ccccc} 1 & \rm{ if } & X_i  & \in & p_\mu \\
0 & \rm{ if } & X_i  & \notin & p_\mu
\end{array}\right. .
\label{Xi_to_pmu}
\eeq
We refer to $P$ as the {\it perfect matching matrix}. This map expresses all chiral fields, {\it including those associated to external edges} in terms of perfect matchings. For the example shown in \fref{dual_quiver}, the perfect matching matrix is displayed in \eref{matching26}.

For BFTs, the master space is toric and its toric diagram is the first polytope we are concerned with. Following the previous discussion, every perfect matching corresponds to a distinct point, i.e.\ a gauged linear sigma model (GLSM) field in the {\it toric diagram of the master space}. In analogy with the planar case, we will often refer to it also as the matching polytope. Relations between these fields are encoded in linear relations between the positions of the corresponding points in the toric diagram. The toric diagram of the master space is given by $P$ which can be row-reduced to simplify its interpretation. For the example given by \fref{dual_quiver}, the master space toric diagram is

{\scriptsize
\begin{equation}
\label{eq:masterspace}
G_{\text{master}}=
\left(
\begin{array}{ccccccccccccccccccccccccc}
 p_1 & p_2 & p_3 & p_4 & p_5 & p_6 & p_7 & p_8 & p_9 & p_{10} & p_{11} & p_{12} & p_{13} & p_{14} & p_{15} & p_{16} & p_{17} & p_{18} & p_{19} & p_{20} & p_{21} & p_{22} & p_{23} & p_{24} & p_{25} \\
 \hline
 1 & 0 & 0 & 0 & 0 & -1 & 0 & 0 & -1 & 0 & -1 & 0 & 0 & -1 & -1 & 0 & 0 & -1 & -1 & 0 & -1 & -1 & 0 & -1 & -1 \\
 0 & 1 & 0 & 0 & 0 & 1 & 0 & 0 & 0 & -1 & 1 & 0 & 0 & 1 & 1 & 0 & 0 & 0 & 0 & -1 & 1 & 0 & -1 & 1 & 0 \\
 0 & 0 & 1 & 0 & 0 & 0 & 0 & 0 & 1 & 1 & 0 & 0 & 0 & 0 & 0 & -1 & -1 & 0 & 0 & 0 & -1 & 1 & 0 & -1 & 1 \\
 0 & 0 & 0 & 1 & 0 & 0 & 0 & 0 & 0 & 0 & 0 & 0 & 0 & 0 & 0 & 1 & 1 & 1 & 1 & 1 & 1 & 0 & 1 & 1 & 0 \\
 0 & 0 & 0 & 0 & 1 & 1 & 0 & 0 & 1 & 0 & 0 & 0 & 0 & 0 & 0 & 0 & 0 & 1 & 1 & 0 & 0 & 0 & 0 & 0 & 0 \\
 0 & 0 & 0 & 0 & 0 & 0 & 1 & 0 & 0 & 1 & 0 & 0 & 0 & 0 & 0 & 0 & 0 & 0 & 0 & 1 & 0 & 0 & 1 & 0 & 0 \\
 0 & 0 & 0 & 0 & 0 & 0 & 0 & 1 & 0 & 0 & 1 & 0 & 0 & 0 & 0 & 1 & 1 & 0 & 0 & 0 & 1 & 0 & 0 & 1 & 0 \\
 0 & 0 & 0 & 0 & 0 & 0 & 0 & 0 & 0 & 0 & 0 & 1 & 0 & 1 & 0 & 0 & -1 & 0 & -1 & 0 & 0 & 1 & -1 & -1 & 0 \\
 0 & 0 & 0 & 0 & 0 & 0 & 0 & 0 & 0 & 0 & 0 & 0 & 1 & 0 & 1 & 0 & 1 & 0 & 1 & 0 & 0 & 0 & 1 & 1 & 1 \\
\end{array}
\right)
\end{equation}
}
from which we see that the toric diagram is 9-dimensional. The coordinates are not all independent however; summing over all 9 rows of $G_{\text{master}}$ we obtain a row of 1's. This is simply the statement that the toric diagram lies on an 8-dimensional hypersurface at unit distance from the origin, which is nothing other than the condition for the toric variety to be CY. We deduce that we have the toric diagram of a 9d CY cone.

The moduli space of the BFT is obtained by projecting the master space onto vanishing D-terms, of which we have one per gauge group or, equivalently, internal face. The moduli space is, once again, a toric CY. The second polytope we are interested in is the {\it toric diagram of the moduli space}. Once again, borrowing the nomenclature from planar graphs, we will often refer to this polytope as the matroid polytope. Perfect matchings correspond to points in this toric diagram, but the map can be many to one. More concretely, perfect matchings which differ by a set of internal faces map to the same point in the toric diagram of the moduli space. For the example above, the toric diagram of the moduli space is given by

{\scriptsize
\begin{equation}
\label{eq:modulispace}
G_{\text{moduli}}=
\left(
\begin{array}{cccc|ccc|cc|c|c|c|c|cc|c|c|ccc|cc|c|c|c}
 p_1 & p_{13} & p_{17} & p_{23} & p_2 & p_{15} & p_{24} & p_3 & p_{25} & p_4 & p_5 & p_6 & p_7 & p_8 & p_{10} & p_9 & p_{11} & p_{12} & p_{16} & p_{20} & p_{14} & p_{21} & p_{18} & p_{19} & p_{22} \\
 \hline
 1 & 1 & 1 & 1 & 0 & 0 & 0 & 0 & 0 & 0 & 0 & -1 & 0 & 0 & 0 & -1 & -1 & 0 & 0 & 0 & -1 & -1 & -1 & 0 & -1 \\
 0 & 0 & 0 & 0 & 1 & 1 & 1 & 0 & 0 & 0 & 0 & 1 & 0 & -1 & -1 & 0 & 0 & -1 & -1 & -1 & 0 & 0 & 0 & 1 & -1 \\
 0 & 0 & 0 & 0 & 0 & 0 & 0 & 1 & 1 & 0 & 0 & 0 & 0 & 1 & 1 & 1 & 1 & 0 & 0 & 0 & 0 & 0 & 0 & 0 & 1 \\
 0 & 0 & 0 & 0 & 0 & 0 & 0 & 0 & 0 & 1 & 0 & 0 & 0 & 0 & 0 & 0 & 0 & 1 & 1 & 1 & 1 & 1 & 1 & 0 & 1 \\
 0 & 0 & 0 & 0 & 0 & 0 & 0 & 0 & 0 & 0 & 1 & 1 & 0 & 0 & 0 & 1 & 0 & 0 & 0 & 0 & 0 & 0 & 1 & 1 & 0 \\
 0 & 0 & 0 & 0 & 0 & 0 & 0 & 0 & 0 & 0 & 0 & 0 & 1 & 1 & 1 & 0 & 1 & 1 & 1 & 1 & 1 & 1 & 0 & -1 & 1 \\
\end{array}
\right)
\end{equation}
}
where we have grouped together the perfect matchings that map to the same point in the toric diagram. Again, all the rows add up to 1, so we see that the toric diagram only lives on a 5-dimensional hyperplane, thus giving us a 6-dimensional toric CY cone.

\bigskip

%===============================================================================
\subsection*{Further Thoughts on the Moduli Space}
%===============================================================================

We would like to use this review as an opportunity to expand on some subtle issues related to external legs in our interpretation of the moduli space.

First, in our analysis the fields associated to external legs are not dynamical, i.e.\ we do not impose vanishing of their F-terms. This has several motivations. Since these fields appear in a single superpotential term, setting their F-terms to zero would set to zero the products of fields they couple to, reducing the moduli space dramatically and potentially making it disappear. Our treatment also matches nicely with matroid polytopes and the boundary measurement, as explained below in \sref{MatroidsandPolytopes} and \sref{sec:BoundMeas}. Furthermore, it beautifully agrees with the subsequent mathematical work on cluster categories for Grassmannians \cite{Baur:2013hwa}. Finally, as discussed in \sref{section_string_embedding}, in those BFTs with a D-brane embedding the fields associated to external legs have a higher dimensional support and are hence naturally non-dynamical from a 4d viewpoint. 

Being non-dynamical, these fields should be regarded as couplings in 4d. However, the second distinctive feature of our treatment of external legs is that we include them as continuous parameters in the moduli space.  We can think about the resulting geometry as a {\it generalized moduli space}, which incorporates all possible values of the corresponding superpotential couplings. The reason why this approach is useful is because the resulting space can be nicely treated in terms of toric geometry. In fact, if desired, it is straightforward to recover the standard moduli space, in which the values of these couplings are fixed, from the generalized one. We just consider slices of the toric generalized moduli space corresponding to setting every external leg $X_e$ to a fixed value $X_{e,0}$. Using \eref{X_from_p}, these constraints can be expressed in terms of perfect matchings as follows:

\beq
X_e=\prod_\mu p_\mu^{P_{e\mu}}=X_{e,0}.
\eeq

\bigskip

%===============================================================================
\subsection{Polytopes from Flows}
%===============================================================================

\label{section_polytopes_from_flows}

The matching and matroid polytopes we introduced in the previous section can alternatively be obtained in terms of flows in a perfect orientation.\footnote{The resulting polytopes are independent of the choice of underlying perfect orientation.} They encode how they can be expressed in terms of natural variables on the graph. 
 
All flows can be specified using a basis of paths which, for graphs on a disk, can be taken as follows:

\begin{itemize}
\item Clockwise loops $w_i$ around each internal face $i$. 
\item Clockwise loops $x_j$ around each external face $j$. These begin and end on boundaries.
\end{itemize}
Notice that the loops $w_i$ and $x_j$ give an orientation to the edges which is not necessarily the same as that defined by the chosen perfect orientation. These loops can be multiplied together to form more complicated loops; when loops share edges with opposite orientations, their effect cancels. 

These loops are not all independent: they satisfy $\prod_{i=1}^{F_e} w_i \prod_{j=1}^{F_i} x_j=1$, with $F_e$ and $F_i$ the numbers of external and internal faces, respectively. It is thus possible to omit one of the faces, which for concreteness we take to be one of the external ones. Expressing flows $\mathfrak{p}_\mu$ as products of loop variables, they are mapped to points in a space of dimension $F-1$, with $F$ the total number of faces, as follows:
\begin{equation}
\label{eq:flowToMaster}
\mathfrak{p}_\mu=\prod_{i=1}^{F_i} w_i^{a_{i,\mu}} \prod_{j=1}^{F_e-1} x_j^{b_{j,\mu}} \ \ \ \ \mapsto \ \ \ \ \begin{array}{c}{\rm \underline{Coordinates}:} \\ (a_{1,\mu},\ldots,a_{F_i,\mu},b_{1,\mu},\ldots,b_{F_e-1,\mu}) \end{array}
\end{equation}
This map is injective: the flows are uniquely specified by the combination of variables $w_i$ and $x_j$. We collect the coordinates into a matrix $G_{\text{matching}}$, where the columns are precisely the coordinates of each $\mathfrak{p}_\mu$, as given by \eref{eq:flowToMaster}. Returning to the example at hand, let us consider the graph in \fref{dual_quiver}.a, and take $p_{\text{ref}}=p_1$. For later convenience, we collect the columns that have the same $b_i$ coordinates. $G_{\text{matching}}$ becomes 

{\scriptsize
\begin{equation}
\label{eq:MatchingPolyForFlows}
\hspace{-.4cm}G_{\text{matching}}=\left(
\begin{array}{r|cccc|ccc|cc|c|c|c|c|cc|c|c|ccc|cc|c|c|c}
  & \mathfrak{p}_1 & \mathfrak{p}_{13} & \mathfrak{p}_{17} & \mathfrak{p}_{23} & \mathfrak{p}_2 & \mathfrak{p}_{15} & \mathfrak{p}_{24} & \mathfrak{p}_3 & \mathfrak{p}_{25} & \mathfrak{p}_4 & \mathfrak{p}_5 & \mathfrak{p}_6 & \mathfrak{p}_7 & \mathfrak{p}_8 & \mathfrak{p}_{10} & \mathfrak{p}_9 & \mathfrak{p}_{11} & \mathfrak{p}_{12} & \mathfrak{p}_{16} & \mathfrak{p}_{20} & \mathfrak{p}_{14} & \mathfrak{p}_{21} & \mathfrak{p}_{18} & \mathfrak{p}_{19} & \mathfrak{p}_{22} \\
  \hline
 a_1 \ & 0 & 1 & 1 & 1 & 0 & 1 & 1 & 0 & 1 & 0 & 0 & 0 & 0 & 0 & 0 & 0 & 0 & 0 & 0 & 0 & 0 & 0 & 0 & 1 & 0 \\
 a_2 \ & 0 & 0 & 1 & 1 & 0 & 0 & 1 & 0 & 0 & 0 & 0 & 0 & 0 & 0 & 0 & 0 & 0 & -1 & 0 & 0 & -1 & 0 & 0 & 1 & -1 \\
 a_3 \ & 0 & 0 & 0 & 1 & 0 & 0 & 0 & 0 & 0 & 0 & 0 & 0 & 0 & -1 & 0 & 0 & -1 & -1 & -1 & 0 & -1 & -1 & 0 & 1 & -1 \\
 \hline
 b_1 \ & 0 & 0 & 0 & 0 & 0 & 0 & 0 & 0 & 0 & 0 & -1 & -1 & -1 & -1 & -1 & -1 & -1 & -1 & -1 & -1 & -1 & -1 & -1 & 0 & -1 \\
 b_2 \ & 0 & 0 & 0 & 0 & 0 & 0 & 0 & 0 & 0 & 0 & 0 & 0 & -1 & -1 & -1 & 0 & -1 & -1 & -1 & -1 & -1 & -1 & 0 & 1 & -1 \\
 b_3 \ & 0 & 0 & 0 & 0 & -1 & -1 & -1 & -1 & -1 & -1 & 0 & -1 & -1 & -1 & -1 & -1 & -2 & -1 & -1 & -1 & -2 & -2 & -1 & 0 & -2 \\
 b_4 \ & 0 & 0 & 0 & 0 & 0 & 0 & 0 & -1 & -1 & -1 & 0 & 0 & 0 & -1 & -1 & -1 & -1 & -1 & -1 & -1 & -1 & -1 & -1 & 0 & -2 \\
 b_5 \ & 0 & 0 & 0 & 0 & 0 & 0 & 0 & 0 & 0 & -1 & 0 & 0 & 0 & 0 & 0 & 0 & 0 & -1 & -1 & -1 & -1 & -1 & -1 & 0 & -1 \\
\end{array}
\right)
\end{equation}
}
It is easy to check that the resulting 8d matching polytope agrees with the one living in the 8d hypersurface of \eref{eq:masterspace}. It can be regarded as the toric diagram of a 9d toric, CY cone.

The previous discussion can be easily extended to graphs on genus $g$ Riemann surfaces with $B$ boundaries, by expanding the basis of loop variables as follows:
\begin{itemize}
\item For surfaces with $g>0$, include oriented fundamental cycles $\alpha_i$ and $\beta_i$, $i=1,\ldots,g$.
\item For surfaces with $B>1$, include paths $\delta_i$ stretching between pairs of boundaries; there should be $B-1$ of these.
\end{itemize}

\bigskip

%===============================================================================
\subsection*{Matroid Polytope from Flows}
%===============================================================================

In many occasions, we are only interested in more limited information about flows. This is for instance the case when identifying non-vanishing entries of the boundary measurement, for which it is only necessary to know whether a flow between two external nodes exists rather than the precise details of the route taken to connect them; this will be further discussed in \sref{sec:BoundMeas}. For planar graphs, this amounts to projecting the matching polytope by only keeping those degrees of freedom associated to external faces.\footnote{Equivalently, we can translate this into degrees of freedom of external legs.} The resulting polytope coincides with the matroid polytope, introduced in the previous section in terms of the BFT moduli space. 

In the approach discussed in this section, we simply obtain the matroid polytope by ignoring the coordinates associated with internal faces and keeping the $b_i$'s in \eref{eq:flowToMaster}. Flows which only differ by internal faces get identified by this projection, which can generally map multiple vertices $\mathfrak{p}_\mu$ to the same lower-dimensional vertex $\pi_j$. For the example treated in the previous subsection, the new vertices $\pi_j$ are simply given by \eref{eq:flowsToMatroid}, where under each vertex we indicate its multiplicity, i.e.\ the number of flows $\mathfrak{p}_\mu$ that identify into each $\pi_j$.

{\footnotesize
\begin{equation}
\label{eq:flowsToMatroid}
G_{\text{matroid}}=\left(
\begin{array}{r|ccccccccccccccc}
  & \pi _1 & \pi _2 & \pi _3 & \pi _4 & \pi _5 & \pi _6 & \pi _7 & \pi _8 & \pi _9 & \pi _{10} & \pi _{11} & \pi _{12} & \pi _{13} & \pi _{14} & \pi _{15} \\
 \hline
 b_1 \ & 0 & 0 & 0 & 0 & -1 & -1 & -1 & -1 & -1 & -1 & -1 & -1 & -1 & 0 & -1 \\
 b_2 \ & 0 & 0 & 0 & 0 & 0 & 0 & -1 & -1 & 0 & -1 & -1 & -1 & 0 & 1 & -1 \\
 b_3 \ & 0 & -1 & -1 & -1 & 0 & -1 & -1 & -1 & -1 & -2 & -1 & -2 & -1 & 0 & -2 \\
 b_4 \ & 0 & 0 & -1 & -1 & 0 & 0 & 0 & -1 & -1 & -1 & -1 & -1 & -1 & 0 & -2 \\
 b_5 \ & 0 & 0 & 0 & -1 & 0 & 0 & 0 & 0 & 0 & 0 & -1 & -1 & -1 & 0 & -1 \\
 \hline
  & \textbf{4} & \textbf{3} & \textbf{2} & \textbf{1} & \textbf{1} & \textbf{1} & \textbf{1} & \textbf{2} & \textbf{1} & \textbf{1} & \textbf{3} & \textbf{2} & \textbf{1} & \textbf{1} & \textbf{1} \\
\end{array}
\right)
\end{equation}
}
Once again, the result obtained from this procedure coincides with the moduli space computation \eref{eq:modulispace}: the 5d polytope defined by \eref{eq:flowsToMatroid} lives on the 5d hypersurface of \eref{eq:modulispace}.

Similar projections are possible for graphs on Riemann surfaces with higher genus and/or multiple boundaries \cite{Franco:2012mm,Franco:2012wv}. In such cases, a different possibility exists for the projections, in precise correspondence with the alternative gauging briefly mentioned in \sref{section_BFTs}.

\bigskip

%===============================================================================
\subsection{Polytopes from Matroids} \label{MatroidsandPolytopes}
%===============================================================================

Matroid theory is the study of abstract dependences. A matroid of rank $k$ on a set $[n]$ is a non-empty collection $\mathcal{M} \in \binom{n}{k}$
of $k$-element subsets $I$ of $[n]$, called bases of the matroid, such that 
they satisfy the exchange axiom:

\medskip

\emph{For any $I,J$ $\in$ $\mathcal{M}$ and $i \in I$, there exists a $j \in J$ such that $(I \, \backslash \, \{ i \}) \cup \{j \} ~ \in \mathcal{M}$}.

\medskip

\noindent For example, for $n=3$ and $k=2$ a matroid could be $\mathcal{M}=\{12,13,23\}$.

This lead us to the final approach to the matroid polytope, which can be defined as the convex hull of the indicator vectors $e_I$ associated to the bases of the matroid, each one defined as $e_I=\sum_{i\in I} e_i$ where $\{e_i \}$ is the standard unit vector basis of $\mathbb{R}^n$. In other words, the matroid polytope is encoded in the matrix

\begin{equation}
(G_{\text{matroid}})_{i\mu}=\left\{ \begin{array}{ccccc} 1 & \rm{ if } & i  & \in & I_\mu \\
0 & \rm{ if } & i  & \notin & I_\mu
\end{array}\right. \quad \Rightarrow \quad G_{\mathcal{M}} = \left(
\begin{array}{c|ccc}
 & \ I_1 \ & \ I_2 \ & \ I_3 \ \\
 \hline
1 \ \ & 1 & 1 & 0 \\
2 \ \ & 1 & 0 & 1 \\
3 \ \ & 0 & 1 & 1 \\
\end{array}
\right)
\label{matroid_polytope}
\end{equation}
where for clarity we also illustrated the example of $\mathcal{M}=\{12,13,23\}$.

Perfect orientations are mapped to the matroid polytope through their source sets. There are $n$ external nodes and $k$ sources, and external nodes $n_i^{(e)}$, $i=1,\ldots,n$ participate in source sets $I_\mu$. Together, the source sets forms the matroid:
\begin{equation}
(G_{\text{matroid}})_{i\mu}=\left\{ \begin{array}{ccccc} 1 & \rm{ if } & n^{(e)}_i  & \in & I_\mu \\
0 & \rm{ if } & n^{(e)}_i  & \notin & I_\mu
\end{array}\right.
\end{equation}

In general, there can be multiple perfect orientations, or equivalently perfect matchings, with the same source set. This will turn out to give exactly the same multiplicities observed in the previous approaches. To illustrate this phenomenon and to compare with the other methods, let us consider the example in \fref{dual_quiver}. The source set of each perfect matching $p_i$, the whole list of which is provided in Appendix \ref{app:PMlist}, is given by

\medskip

{\small
\begin{eqnarray}
\begin{array}{cccccccccccccc}
 p_1 & p_2 & p_3 & p_4 & p_5 & p_6 & p_7 & p_8 & p_9 & p_{10} & p_{11} & p_{12} & p_{13} & p_{14} \\
 \downarrow & \downarrow & \downarrow & \downarrow & \downarrow &\downarrow & \downarrow & \downarrow & \downarrow & \downarrow & \downarrow & \downarrow &\downarrow & \downarrow \\
 \{2,6\} & \{3,6\} & \{4,6\} & \{5,6\} & \{1,2\} & \{1,3\} & \{2,3\} & \{2,4\} & \{1,4\} & \{2,4\} & \{3,4\} & \{2,5\} & \{2,6\} & \{3,5\} \\
\end{array} \nonumber \\
\begin{array}{ccccccccccc}
p_{15} & p_{16} & p_{17} & p_{18} & p_{19} & p_{20} & p_{21} & p_{22} & p_{23} & p_{24} & p_{25} \\
 \downarrow & \downarrow & \downarrow & \downarrow & \downarrow & \downarrow & \downarrow & \downarrow & \downarrow & \downarrow & \downarrow \\
 \{3,6\} & \{2,5\} & \{2,6\} & \{1,5\} & \{1,6\} & \{2,5\} & \{3,5\} & \{4,5\} & \{2,6\} & \{3,6\} & \{4,6\} \\
\end{array} \quad \quad \quad \quad \;
\end{eqnarray}
}

\medskip

Mapping each source set to the corresponding matroid polytope coordinate, we get a matroid polytope summarized by the following matrix
{\small
\begin{equation}
\left(
\begin{array}{cccc|ccc|cc|c|c|c|c|cc|c|c|ccc|cc|c|c|c}
 p_1 & p_{13} & p_{17} & p_{23} & p_2 & p_{15} & p_{24} & p_3 & p_{25} & p_4 & p_5 & p_6 & p_7 & p_8 & p_{10} & p_9 & p_{11} & p_{12} & p_{16} & p_{20} & p_{14} & p_{21} & p_{18} & p_{19} & p_{22} \\
 \hline
 0 & 0 & 0 & 0 & 0 & 0 & 0 & 0 & 0 & 0 & 1 & 1 & 0 & 0 & 0 & 1 & 0 & 0 & 0 & 0 & 0 & 0 & 1 & 1 & 0 \\
 1 & 1 & 1 & 1 & 0 & 0 & 0 & 0 & 0 & 0 & 1 & 0 & 1 & 1 & 1 & 0 & 0 & 1 & 1 & 1 & 0 & 0 & 0 & 0 & 0 \\
 0 & 0 & 0 & 0 & 1 & 1 & 1 & 0 & 0 & 0 & 0 & 1 & 1 & 0 & 0 & 0 & 1 & 0 & 0 & 0 & 1 & 1 & 0 & 0 & 0 \\
 0 & 0 & 0 & 0 & 0 & 0 & 0 & 1 & 1 & 0 & 0 & 0 & 0 & 1 & 1 & 1 & 1 & 0 & 0 & 0 & 0 & 0 & 0 & 0 & 1 \\
 0 & 0 & 0 & 0 & 0 & 0 & 0 & 0 & 0 & 1 & 0 & 0 & 0 & 0 & 0 & 0 & 0 & 1 & 1 & 1 & 1 & 1 & 1 & 0 & 1 \\
 1 & 1 & 1 & 1 & 1 & 1 & 1 & 1 & 1 & 1 & 0 & 0 & 0 & 0 & 0 & 0 & 0 & 0 & 0 & 0 & 0 & 0 & 0 & 1 & 0 \\
\end{array}
\right) \label{eq:PerfOrientandMatroidPolytope}
\end{equation}
}

This is equivalent to the toric diagram of the moduli space \eref{eq:modulispace}. As in \eref{eq:modulispace}, one row is redundant: \eref{eq:PerfOrientandMatroidPolytope} specifies a 5d polytope, living on a hypersurface at a distance from the origin, thus describing a 6d toric CY cone.

The methods discussed in \sref{section_master_and_moduli_spaces} and \sref{section_polytopes_from_flows} apply to general graphs, including those {\it without boundaries}. The one based on matroids is, however, tied to the existence of external nodes and hence boundaries. It would be interesting to explore whether the notion of matroids can be generalized to deal with arbitrary bipartite graphs.

\bigskip

%===============================================================================
\section{Bipartite Graphs and the Grassmannian}
%===============================================================================

Bipartite graphs are also intimately related to the Grassmannian. The BFT interpretation of bipartite graphs provides useful tools for investigating the Grassmannian, as will be discussed in \sref{sec:BFTtools}. In this section we first review the definition of the Grassmannian and then we describe the map between bipartite graphs and elements in the Grassmannian, known as the {\it boundary measurement}. See \cite{2012arXiv1210.5433T,2006math09764P,2007arXiv0706.2501P,Postnikovlectures} for a detailed exposition of these ideas.

\bigskip

%===============================================================================
\subsection{The Grassmannian and \pl Coordinates}
%===============================================================================

The real Grassmannian $Gr_{k,n}(\mathbb{R})$ is the space of 
$k$-dimensional planes in $n$-dimension passing through the origin. A $k$-plane in $n$ dimensions is specified by $k$ $n$-dimensional vectors, which can be arranged into a $k \times n$ matrix $C$. The plane spanned by these vectors
is invariant under the action of the $GL(k)$ on this matrix. The Grassmannian $Gr_{k,n}(\mathbb{R})$ is thus the space of $k \times n$ matrices with real entries modulo the action of $GL(k)$. Its dimension is $k(n-k)$.

The $GL(k)$ invariant parameterization of the Grassmannian \Gr$(\mathbb{R})$ is given by ratios of all maximal $k \times k$ minors of the matrix $C$. These minors are known as {\it \pl coordinates} and we denote them $\Delta_I$, where $I$ indicates the set of $k$ columns involved in the determinant. \pl coordinates are invariant under $SL(k)$ and rescale with a common factor under $GL(k)$. 

Since the $\binom{n}{k} $ minors of a $k \times n$ matrix are not independent, \pl coordinates obey the so-called \pl relations:

\begin{equation}
\sum_{i=1}^{k+1} (-1)^{i-1} \Delta_{I_1 \cup \,a_i} \, \Delta_{I_2\, \backslash \,a_i} = 0 ,
\end{equation}
Here $I_1$ is any $(k-1)$-element subset of $[n]$, $I_2$ is any $(k+1)$-element subset of $[n]$ and $a_i$ is the $i$th element of $I_2$. In each term of the sum, $a_i$ is deleted from $I_2$ and appended to the right of $I_1$. \pl coordinates together with the relations they satisfy
define the embedding 
\Gr$(\mathbb{R})$ $\hookrightarrow \mathbb{RP}^{\binom{n}{k}-1}$.
 
The positive Grassmannian, also denoted $Gr_{k,n}^{\geq 0}(\mathbb{R})$, is the restriction of Grassmannian to non-negative \pl coordinates.

\bigskip

%===============================================================================
\subsection{The Matroid Stratification of the Grassmannian}
%===============================================================================

The theory of matroids leads to a useful decomposition of the Grassmannian, its  {\it matroid stratification}. Let $\mathcal{M} \in \binom{n}{k}$ be a matroid, the corresponding matroid stratum of the Grassmannian is defined as
$$
S_{\mathcal{M}}=\{ C \in Gr_{k,n} | ~\Delta_I \neq 0 ~\text{if and only if} ~ I \in \mathcal{M} \}
$$
Each stratum is thus specified by the set of vanishing and non-vanishing \pl coordinates.

\bigskip

%===============================================================================
\subsection{The Boundary Measurement} \label{sec:BoundMeas}
%===============================================================================

The {\it boundary measurement} is a map from edge weights in a bipartite graph to the Grassmannian. It was originally introduced for planar graphs by Postnikov in \cite{2006math09764P}. The boundary measurement was later extended to bipartite graphs on the annulus in \cite{2009arXiv0901.0020G}, and to multiple boundaries in \cite{Franco:2013nwa}.

\bigskip

%===============================================================================
\subsubsection{Planar Graphs} \label{sec:planarBoundMeas}
%===============================================================================

Let us focus on planar bipartite graphs with $n$ external nodes such that any perfect orientation has $k$ sources. To parametrize paths, we introduce variables $\alpha_i$, which are {\it oriented edge weights}. We adopt the convention in which the orientation goes from white to black nodes. If the edge is traversed in the opposite direction, we associate to it a weight $\alpha_i^{-1}$.

The boundary measurement is a $k\times n$ matrix $C$ constructed as follows:

\medskip

\begin{itemize}
\item Label the external nodes in clockwise order.
\item Select a perfect orientation (equivalently a reference perfect matching). The choice of the perfect orientation identifies $k$ sources.
\item The entry $C_{ij}$ is given by the `sum' of weights of the oriented paths in the perfect orientation connecting source $i$ to external node $j$. In fact some relative signs between the contributions of different paths need to be included: for every loop in a given path, we introduce an additional $(-1)$ factor. We also introduce an overall $(-1)^{s(i,j)}$ to each entry,  where $s(i,j)$ counts the number of sources strictly between $i$ and $j$, neglecting periodicity.
\end{itemize}

\medskip

Trivial paths going from a source to itself correspond to entries equal to $1$. If there is no oriented path between a source and an external node, the corresponding entry vanishes.

An important implication of the delicate prescription concerning the signs is that it guarantees that, for real and non-negative edge weights, $C$ has positive minors. The boundary measurement thus provides a map to the positive Grassmannian $Gr_{k,n}^{\geq 0}(\mathbb{R})$. This property is unique of planar graphs.

The paths contributing to entries of $C$ can be identified with {\it single component} flows. Moreover, another non-trivial consequence of the sign assignments explained above is that \pl coordinates, i.e.\ the $k\times k$ minors, can be expressed as sum of flows. This remarkable property is preserved by the generalized boundary measurement definitions of \cite{2009arXiv0901.0020G} and \cite{Franco:2013nwa}. This, in turn, implies that there is a map between \pl coordinates and perfect matchings, as we discuss below. This map is independent of the choice of perfect orientation, equivalently of reference perfect matching.

Let us see these ideas at work for the example in \fref{dual_quiver}, for which the perfect matchings are listed in Appendix \ref{app:PMlist}. Let us consider the perfect orientation in \fref{pms_orientations_flows}.b, which corresponds to the perfect matching $p_5$.

%===============================================================================
\begin{figure}[h]
\begin{center}
\includegraphics[width=13.5cm]{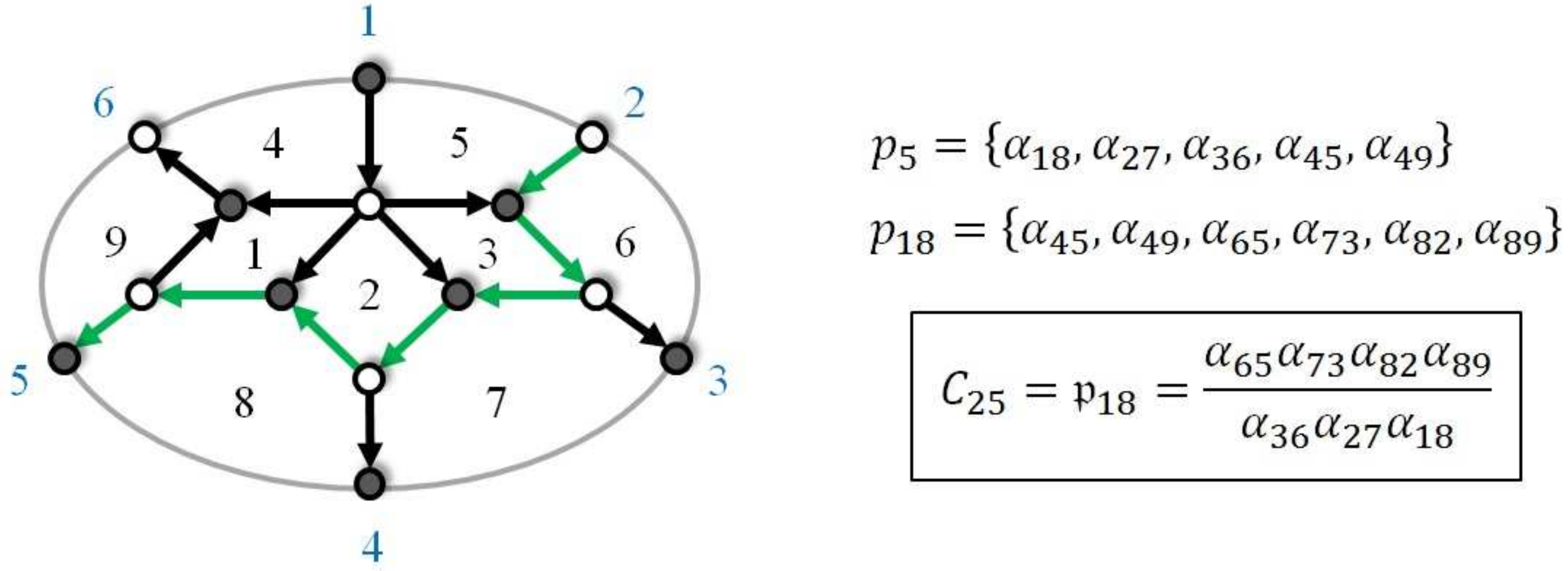}
\caption{Example of $C_{25}$ entry in the boundary measurement. There is a single oriented path connecting nodes 2 and 5, i.e.\ the flow $\mathfrak{p}_{18}$.}
\label{boundary_measurement}
\end{center}
\end{figure}
%===============================================================================

\fref{boundary_measurement} shows how to obtain the $C_{25}$ entry using the boundary measurement. The full boundary measurement for this example, written in terms of flows, is given by 
\be
C=\left(
\begin{array}{c|cccccc}
& 1 & 2 & 3 & 4 & 5 & 6   \\
\hline
1 \ \ & \ 1 \ & \ 0 \ &\ \ -\mathfrak{p}_7 \ \ &\ \ - {\mathfrak{p}_8} - {\mathfrak{p}_{10}} \ \ &
\ \ - {\mathfrak{p}_{12}}- {\mathfrak{p}_{16}} - {\mathfrak{p}_{20}} \ \ &
\ \ - {\mathfrak{p}_1} - {\mathfrak{p}_{13}} - {\mathfrak{p}_{17}} \ \ 
\\
2\ \ & 0 & 1 &  {\mathfrak{p}_6}  &   {\mathfrak{p}_9}  &   {\mathfrak{p}_{18}}  &   {\mathfrak{p}_{19}} 
\end{array}
\right)
\ee
We see that this graph, which will be one of our main examples throughout this article, corresponds to a top-dimensional cell of $Gr_{2,6}$. The \pl coordinates become

\beq
\begin{array}{rclcrclcrclcrcl}
\Delta_{12} & = & 1 & \ \ \ & \Delta_{13} & = & {\mathfrak{p}_6} & \ \ \ & \Delta_{14} & = & {\mathfrak{p}_9} & \ \ \ & \Delta_{15} & = & {\mathfrak{p}_{18}} \\
\Delta_{16} & = & {\mathfrak{p}_{19}} & &
\Delta_{23} & = & {\mathfrak{p}_7} & & 
\Delta_{24} & = & {\mathfrak{p}_8} + {\mathfrak{p}_{10}} & &
\Delta_{25} & = & {\mathfrak{p}_{12}}+ {\mathfrak{p}_{16}} + {\mathfrak{p}_{20}} \\
\Delta_{26} & = & {\mathfrak{p}_1} + {\mathfrak{p}_{13}} + {\mathfrak{p}_{17}} & &
\Delta_{34} & = & {\mathfrak{p}_{11}}  & &
\Delta_{35} & = & {\mathfrak{p}_{14}} + {\mathfrak{p}_{21}} & & 
\Delta_{36} & = & {\mathfrak{p}_{2}} + {\mathfrak{p}_{15}} + {\mathfrak{p}_{24}} 
\\
\Delta_{45} & = & {\mathfrak{p}_{22}} + {\mathfrak{p}_{15}} + {\mathfrak{p}_{24}} & &
\Delta_{46} & = & {\mathfrak{p}_{3}} + {\mathfrak{p}_{25}} & &
\Delta_{56} & = & {\mathfrak{p}_{4}} 
\end{array}
\eeq
where the subscripts indicates the columns involved in the minors. They are all non-negative for non-negative edge weights.

As mentioned earlier, the sign assignments are crucial for the simplifications that turn \pl coordinates into sums of flows.

\bigskip

%===============================================================================
\subsection{\pl Coordinates and Perfect Matchings}
%===============================================================================

We have just seen that there is a precise map between \pl coordinates and flows. Since perfect matchings are in bijection with flows as described in \sref{sec:BipaTec}, this map can be translated into a map between \pl coordinates and perfect matchings. It is important to notice that a single \pl coordinate can correspond to several perfect matchings.

The map can be easily implemented as follows. Every perfect matching defines a perfect orientation with source set $I$, which is identified with the set of columns in $C$ associated to a \pl coordinate. Thus, this perfect matching contributes to the \pl coordinate $\Delta_I$. This prescription extends to non-planar graphs \cite{Franco:2013nwa}. In summary:
\beq
\label{eq:ImportantRelation}
\{ \text{Perfect matching} \rightarrow \text{Perfect orientation} \rightarrow \text{Source set } I \} \ \  \Longleftrightarrow \ \  \Delta_I .
\eeq

\bigskip

%===============================================================================
\subsection{Boundary Measurement Beyond Planar Graphs} \label{sec:nonplanar}
%===============================================================================

Non-planar graphs also have a map to elements of the Grassmannian. In this section we demonstrate the subtleties that need to be addressed by a non-planar boundary measurement, and the prescription to implement them introduced in \cite{Franco:2013nwa}. We shall restrict our discussion to graphs with genus zero and multiple boundaries. For cases on the annulus, a well-defined map to the Grassmannian already exists \cite{2009arXiv0901.0020G}; the one presented here reduces to the known cases on the annulus and the disk in the cases with two or one boundary, respectively, and can be seen as a generalization of them.

As in the planar boundary measurement, for a given perfect orientation, the matrix entries $C_{ij}$ of the element of the Grassmannian are composed of paths connecting the $k$ sources to the $n$ external nodes. Given the relation \eref{eq:ImportantRelation} between perfect orientations and matroid elements, and hence minors of the Grassmannian, we insist that minors of $C$ be expressed as `sums' of flows. This typically requires a delicate assignation of signs in the matrix entries $C_{ij}$.

There are two principal sources of difficulty:
\begin{itemize}
\item The ordering of external nodes determines the position of the corresponding columns in $C$, thus affecting the signs associated with minors involving that column.
\item Each loop gives a $(-1)$ sign to a given flow in $C_{ij}$. We will need a more general prescription for counting loops.
\end{itemize}
For compatibility with the known planar boundary measurement, we will also need to keep the sign $(-1)^{s(i,j)}$ introduced in \sref{sec:planarBoundMeas} that is given to a matrix entry $C_{ij}$. 

\begin{figure}[h]
\begin{center}
\includegraphics[scale=0.3]{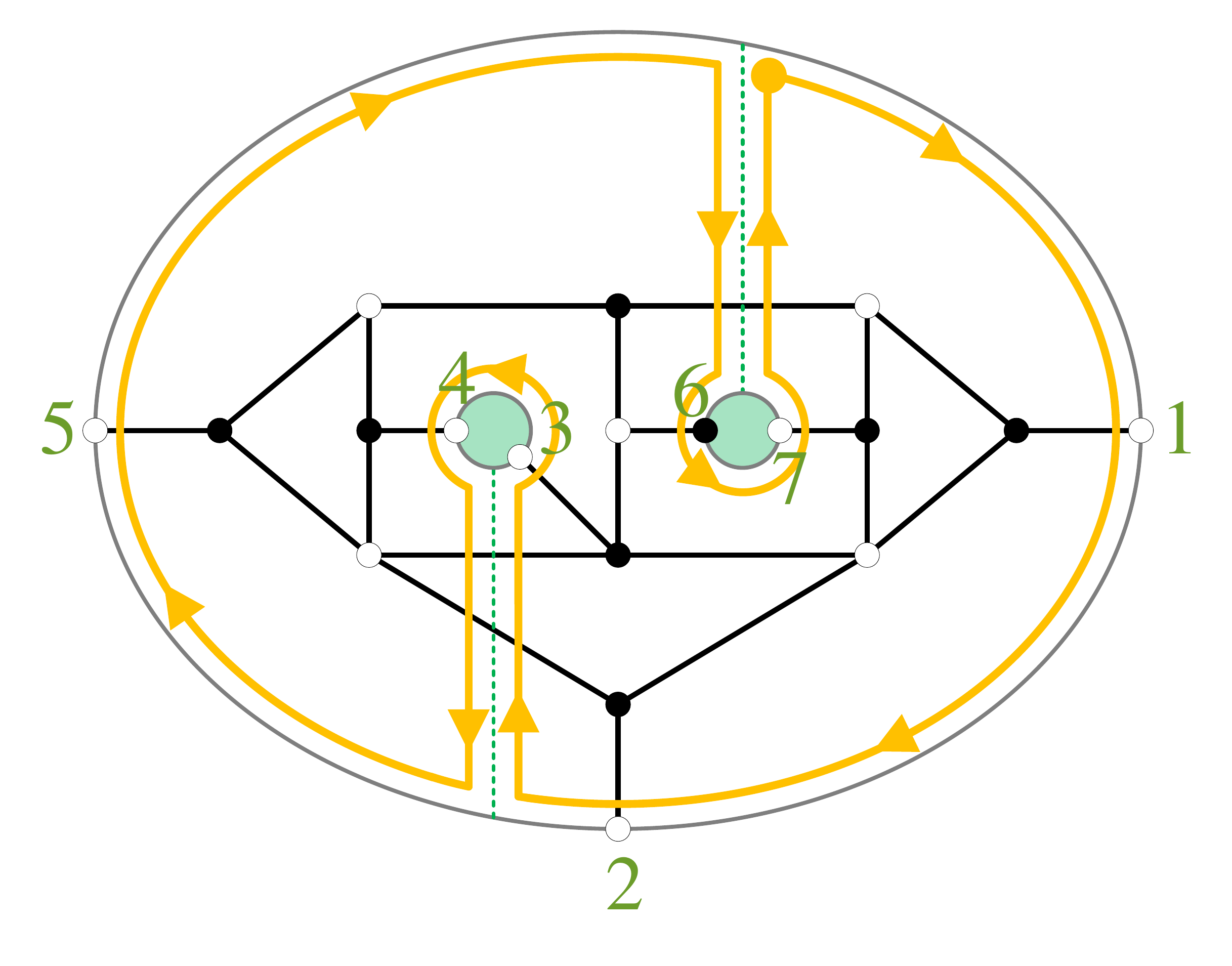}
\caption{Labeling of external nodes by following boundaries and cuts. The cuts are represented by green dotted lines.}
\label{fig:3BexternalLabels}
\end{center}
\end{figure}

To address both issues, we introduce \textit{cuts} between boundaries. These cuts might cross over some of the edges of the graph. The ordering prescription for the external nodes is fixed by creating a path along the cuts and boundaries, in a way which reminds of the computation of residues in complex analysis: we start at an arbitrary point on one of the boundaries, and follow the boundary until reaching a cut. Then we follow the cut to the next boundary, follow the boundary to the next cut, and so on, until reaching the original starting point. This should be done without ever crossing over any cuts or boundaries. An example of this, taken from \cite{Franco:2013nwa}, is given in \fref{fig:3BexternalLabels}. External nodes are labeled according to the order in which they appear along the path.

%===============================================================================
%===============================================================================

\begin{figure}[h]
\begin{center}
\includegraphics[scale=0.6]{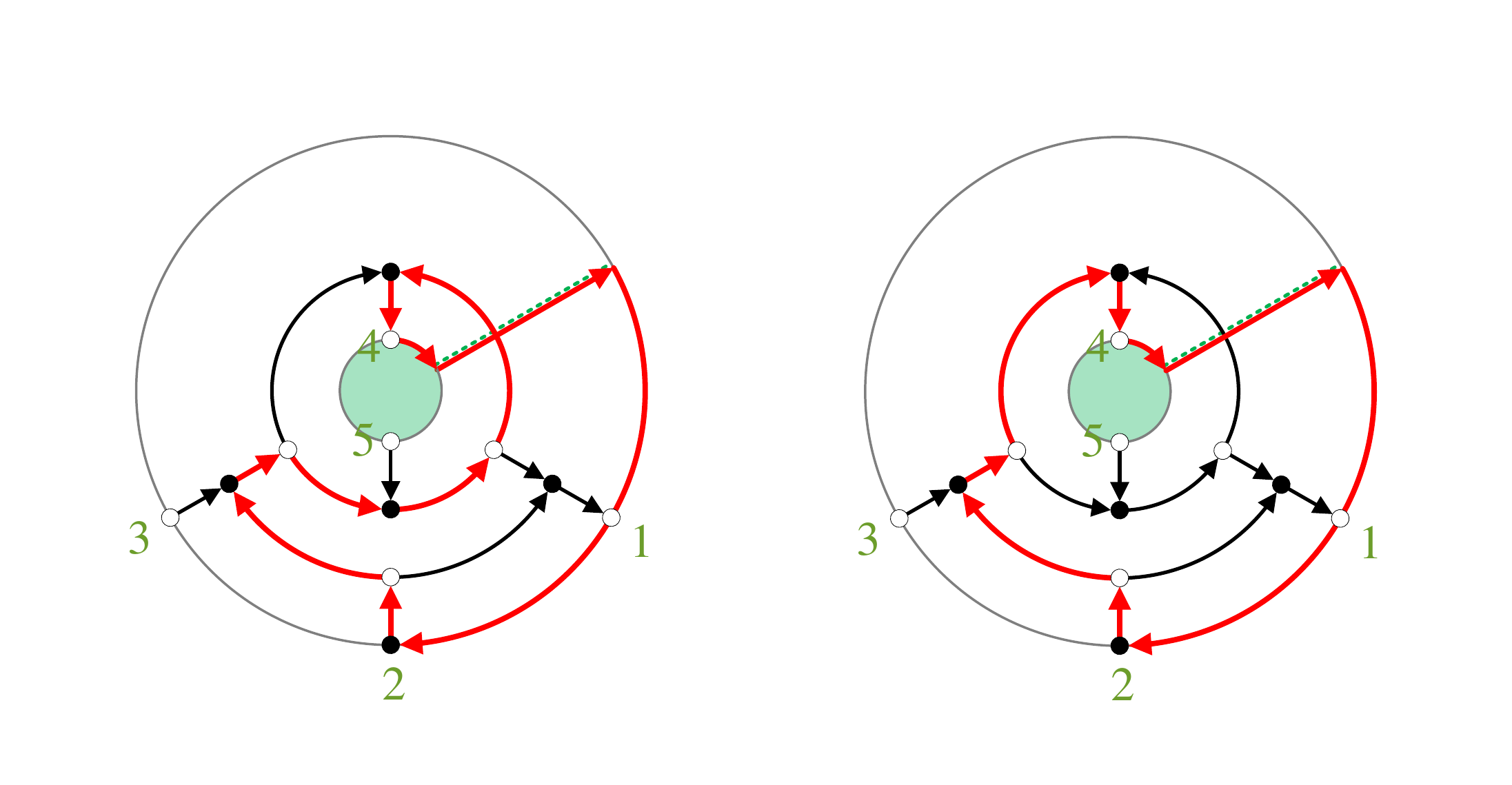}
\caption{The path is closed using cuts and boundaries. The example on the left is composed of two loops and gives a sign $(-1)$ to this contribution in $C_{24}$. The example on the right is a single loop, and gives no additional signs to this contribution in $C_{24}$.}
\label{fig:2Bloopflows}
\end{center}
\end{figure}

Matrix entries $C_{ij}$ are composed of paths from source $i$ to node $j$. To count loops for each path, we first \textit{close the path} by starting from the sink and following a succession of boundaries and cuts in order to get from the sink to the source. With the closed path that is formed in this way, we simply identify all loops that are formed. This includes those loops introduced in \sref{sec:planarBoundMeas}, but generally includes new loops, in particular when the path uses cuts that cross over edges used by the path. The sign assigned to each path is $(-1)^{L-1}$, where $L$ is the number of independent loops in the closed path. Two examples of this are given in \fref{fig:2Bloopflows}.

%===============================================================================
%===============================================================================

In summary, the construction of the non-planar boundary measurement is the following:
\begin{itemize}
\item Choose cuts. Label the $n$ external nodes according to the prescription illustrated in \fref{fig:3BexternalLabels}.
\item Choose a perfect orientation, which determines a source set of with $k$ external nodes.
\item Construct a $k \times n$ matrix $C$, with entries $C_{ij}$ equal to the sum of connected oriented paths from source $i$ to external node $j$, as in \sref{sec:planarBoundMeas}. So far we have not introduced any signs.
\item For each contribution to $C_{ij}$, close the path using the boundaries and cuts. Determine the number of loops $L$ and assign a $(-1)^{L-1}$ to this contribution in $C_{ij}$.
\item Give an additional overall sign $(-1)^{s(i,j)}$ to matrix entries $C_{ij}$, as in \sref{sec:planarBoundMeas}.
\end{itemize}

It is important to note that this boundary measurement reduces to the known cases of the disk \cite{2006math09764P} and the annulus \cite{2009arXiv0901.0020G}, and does not depend on the choice of cuts. Contrary to the planar case, the \pl coordinates are no longer positive definite, given positive oriented edge weights. However, a remarkable property of the boundary measurement we have introduced is that the minors of $C$ can be expressed as simple sums of flows, in accordance with the map between perfect matchings and matroid bases, as in \sref{eq:ImportantRelation}.

\bigskip

%===============================================================================
\section{ Bipartite Graphs as On-Shell Diagrams}
%===============================================================================

Remarkably, bipartite graphs recently played a prominent role in the context of a different physical problem. In \cite{ArkaniHamed:2012nw}, the computation of scattering amplitudes in 4d $\mathcal{N}=4$ SYM has been reformulated in terms of {\it on-shell diagrams}.\footnote{These ideas have been extended to the 3d ABJM theory \cite{2008JHEP...10..091A} in \cite{2014JHEP...02..104H,Huang:2014xza}.} The new approach makes all symmetries of the theory manifest and sheds new light on previous results \cite{ArkaniHamed:2012nw,ArkaniHamed:2010kv,Britto:2004ap,Britto:2005fq}.

On-shell diagrams are in fact bipartite graphs with boundaries, constructed by attaching MHV and $\overline{\mbox{MHV}}$ 3-point amplitudes.\footnote{It is possible to consider non-bipartite on-shell diagrams, but they can be turned into bipartite ones via a well-defined prescription \cite{Franco:2012mm}.} These diagrams connect scattering amplitudes with the Grassmannian, by means of the boundary measurement map discussed in \sref{sec:BoundMeas}. In this case, $n$ corresponds to total number of scattered gluons and $k$ to number of those with negative helicity. More generally, this gluon scattering amplitude can be viewed as a component in an object containing the scattering of all particles related to them by $\mathcal{N}=4$ SUSY.

For planar graphs, the positroid stratification \cite{2006math09764P}, i.e.\ the boundary structure, of the cell in the positive Grassmannian associated to a given graph encodes the singularity structure of the corresponding scattering integrand.

In the following we will show that the BFT interpretation of bipartite graphs provides powerful tools for investigating their applications in the context of the Grassmannian and on-shell diagrams. When gluing 3-point amplitudes, it is convenient to introduce a gauge redundancy at each node. This gives rise to a $U(1)$ gauge theory on the graph which, for some questions, is equivalent to an Abelian BFT.

\bigskip

%===============================================================================
\section{BFTs as Tools} \label{sec:BFTtools}
%===============================================================================

As explained, one of the reasons that make BFTs attractive is that, as a result of their rather constrained structure, interesting questions such as the computation of their master and moduli spaces become trivial and combinatorial. A different motivation for their study is that some of these problems can shed light on seemingly different, but ultimately {\it equivalent}, questions arising on other systems associated to the same underlying graphs. The purpose of this section is to present two such applications of BFTs: {\it graph equivalence} and the {\it stratification of the Grassmannian}. 

\bigskip

%===============================================================================
\subsection{Graph Equivalence}
%===============================================================================

Graph equivalence is a crucial notion when interpreting bipartite graphs as on-shell diagrams, because equivalent graphs give rise to the same leading singularities \cite{ArkaniHamed:2012nw}. More generally, the matrices obtained via the boundary measurement defined in \sref{sec:BoundMeas}, which were extended to non-planar graphs in \sref{sec:nonplanar}, cover exactly the same region of the Grassmannian. 

Some definitions of graph equivalence are, at least in their current form, exclusive to planar graphs. For example, two planar graphs are equivalent if they give rise to the same {\it permutation}, whose determination in turn requires invoking rather sophisticated constructions such as zig-zag paths \cite{Franco:2012mm,2006math09764P,ArkaniHamed:2012nw}. Two graphs are also equivalent if they can be connected by the graph transformations $a$ to $c$ in \fref{graphical_dynamics}. Remarkably, graph equivalence can be rephrased as follows: two graphs are equivalent if the associated BFTs have the same moduli space\cite{Franco:2012mm}.\footnote{In fact this definition is more general than the one based on graph transformations we just gave.  Sometimes the connection between two equivalent graphs might also involve operation $d$ from \fref{graphical_dynamics}, although this operation does not always lead to equivalent graphs.} This definition in terms of an auxiliary BFT is not only computationally useful, but also applies to non-planar graphs, hence generalizing the notion to them.

\begin{figure}[h]
\begin{center}
\includegraphics[scale=0.6]{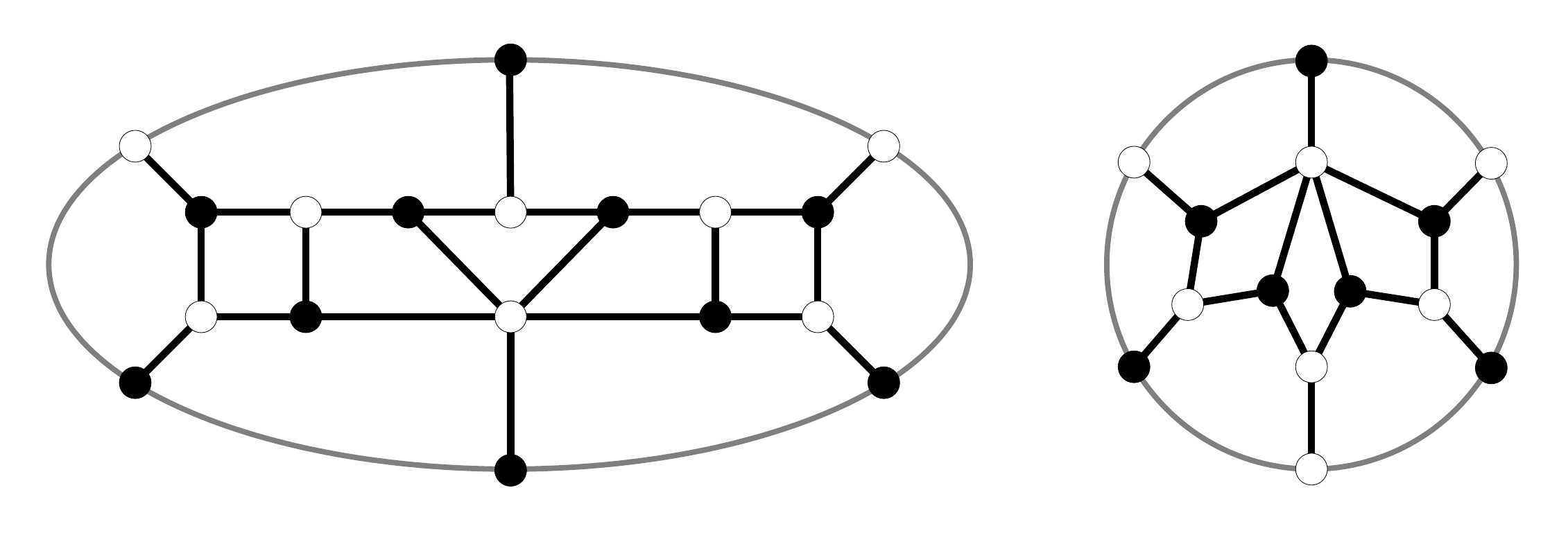}
\caption{Two equivalent graphs. There exists a sequence of graphical operations that turns the graph on the left into the graph on the right, however it is not necessary to find the sequence: it is sufficient to study the moduli space of the two graphs to determine their equivalence.}
\label{fig:graphequiv}
\end{center}
\end{figure}

To illustrate the utility of the BFT perspective on graph equivalence, we provide in \fref{fig:graphequiv} an example of two equivalent graphs.

%===============================================================================
%===============================================================================

The power of BFTs is that it is not necessary to explicitly determine a sequence of operations connecting the two graphs in order to show that they are equivalent, the computation of the BFT moduli space is a sufficient diagnostic for determining the equivalence. Explicitly computing the moduli space for the two graphs at hand, we obtain:

{\small
\begin{equation}
G_{\text{left}}=\left(
\begin{array}{ccccccccccccccc}
 0 & 0 & 0 & 0 & 1 & 1 & 0 & 0 & 1 & 0 & 0 & 0 & 1 & 1 & 0 \\
 0 & 0 & 0 & 0 & 1 & 1 & 1 & 1 & 1 & 1 & 1 & 1 & 1 & 0 & 1 \\
 0 & 1 & 0 & 0 & 0 & 1 & 1 & 0 & 0 & 1 & 0 & 1 & 0 & 0 & 0 \\
 0 & 1 & 1 & 1 & 0 & 1 & 0 & 0 & 1 & 1 & 0 & 1 & 1 & 1 & 1 \\
 0 & 0 & 1 & 0 & 0 & 0 & 0 & 1 & 1 & 1 & 0 & 0 & 0 & 0 & 1 \\
 0 & 0 & 0 & 1 & 0 & 0 & 0 & 0 & 0 & 0 & 1 & 1 & 1 & 0 & 1 \\
 \hline
 \textbf{12} & \textbf{7} & \textbf{2} & \textbf{1} & \textbf{3} & \textbf{2} & \textbf{1} & \textbf{2} & \textbf{1} & \textbf{1} & \textbf{7} & \textbf{4} & \textbf{2} & \textbf{3} & \textbf{1} \\
\end{array}
\right) \quad G_{\text{right}}=\left(
\begin{array}{ccccccccccccccc}
 0 & 0 & 0 & 0 & 1 & 1 & 0 & 0 & 1 & 0 & 0 & 0 & 1 & 1 & 0 \\
 0 & 0 & 0 & 0 & 1 & 1 & 1 & 1 & 1 & 1 & 1 & 1 & 1 & 0 & 1 \\
 0 & 1 & 0 & 0 & 0 & 1 & 1 & 0 & 0 & 1 & 0 & 1 & 0 & 0 & 0 \\
 0 & 1 & 1 & 1 & 0 & 1 & 0 & 0 & 1 & 1 & 0 & 1 & 1 & 1 & 1 \\
 0 & 0 & 1 & 0 & 0 & 0 & 0 & 1 & 1 & 1 & 0 & 0 & 0 & 0 & 1 \\
 0 & 0 & 0 & 1 & 0 & 0 & 0 & 0 & 0 & 0 & 1 & 1 & 1 & 0 & 1 \\
 \hline
 \textbf{4} & \textbf{3} & \textbf{2} & \textbf{1} & \textbf{1} & \textbf{1} & \textbf{1} & \textbf{2} & \textbf{1} & \textbf{1} & \textbf{3} & \textbf{2} & \textbf{1} & \textbf{1} & \textbf{1} \\
\end{array}
\right)
\end{equation}
}

The moduli spaces of the two BFTs agree, modulo multiplicities of the points, showing that the underlying graphs are equivalent.

\bigskip

%===============================================================================
\subsection{Combinatorial Stratification of the Grassmannian} \label{sec:CombStrat}
%===============================================================================

In \sref{3polytopeRoutes} we described two polytopes associated to a bipartite diagram, the matching and matroid polytopes. These may be interpreted as the toric diagrams for the master and moduli spaces of the BFT associated with the bipartite graph. The matching polytope contains the full information on perfect matchings. The matroid polytope contains the information on which matroid bases are present for the Grassmannian element associated to the bipartite graph, i.e.\ which \pl coordinates are non-zero. The agreement between the matroid polytope and the toric diagram of the moduli space is an important feature which is respected by the non-planar generalization of the boundary measurement, as discussed in detail in \cite{Franco:2013nwa}.

Given a region or cell in the Grassmannian, we are interested in a detailed geometric characterization of it. In this section we discuss the {\it combinatorial stratification} introduced in \cite{Franco:2013nwa}. This stratification coincides with the {\it positroid stratification} in the case of planar graphs, and provides a partial {\it matroid stratification} for general graphs. The combinatorial stratification is directly related to the singularity structure of the corresponding on-shell diagram. This can be obtained by stratifying the Grassmannian element in lower dimensional components, by successively turning off \pl coordinates while taking into account the \pl relations.

Focusing on the planar case, \cite{2006math09764P,ArkaniHamed:2012nw} showed  that this stratification can be obtained from the bipartite graphs by removing so-called \emph{removable edges}, i.e.\ those edges that after being deleted yield a \textit{reduced} graph, which is a graph with the fewest possible number of faces given a matroid polytope.\footnote{Reduced graphs are only determined up to graph equivalence and are hence not uniquely defined.} This procedure can be efficiently implemented with the tools presented in \cite{Franco:2013nwa}, which also contains a more detailed discussion on this procedure.

We shall now show that the combinatorial stratification of the Grassmannian can be obtained in a different but very efficient way, which makes heavy use of the geometry already associated to the graphs \cite{Franco:2013nwa}. This alternative way of stratifying the Grassmannian element is computationally very powerful, and never makes explicit use of \pl coordinates, removable edges or reducibility. It is deeply motivated by thinking of graphs and geometry in terms of BFTs.

\bigskip

%===============================================================================
\subsubsection{Face Poset of the Matching Polytope}
%===============================================================================

The combinatorial stratification is implemented in two steps. The first one consists in constructing the face lattice of the matching polytope, i.e.\ the poset encoding its geometrical boundaries. The first level of the poset is the matching polytope itself, of dimension $d_{matching}$. The second level contains the collection of faces of dimension $d_{matching}-1$. The third level is obtained by taking the $(d_{matching}-2)$-dimensional boundaries of each of the $(d_{matching}-1)$-dimensional faces. The subsequent levels are obtained in the same way, until reaching the last level, of dimension zero, which contains the vertices of the matching polytope.

There are several different methods available for computing the face lattice of polytopes, see e.g.\ \cite{polymake}. We shall here use a method that directly utilizes the connection between the matching polytope and the bipartite graph, based on ideas from \cite{2007arXiv0706.2501P}. Each face of the matching polytope can be obtained by iteratively removing an equivalence class of edges, where two edges are considered equivalent if they participate in the same set of perfect matchings. Edges can become equivalent {\it after} removing other edges and the associated perfect matchings. The effect this operation has on the perfect matching matrix $P$ is to remove a set of identical rows, and those columns in which the rows had a 1. Successively repeating this procedure for each of the subgraphs, produced by removing a specific equivalence class of edges, gives the face lattice of the matching polytope.

\begin{figure}[h]
\begin{center}
\includegraphics[scale=0.45]{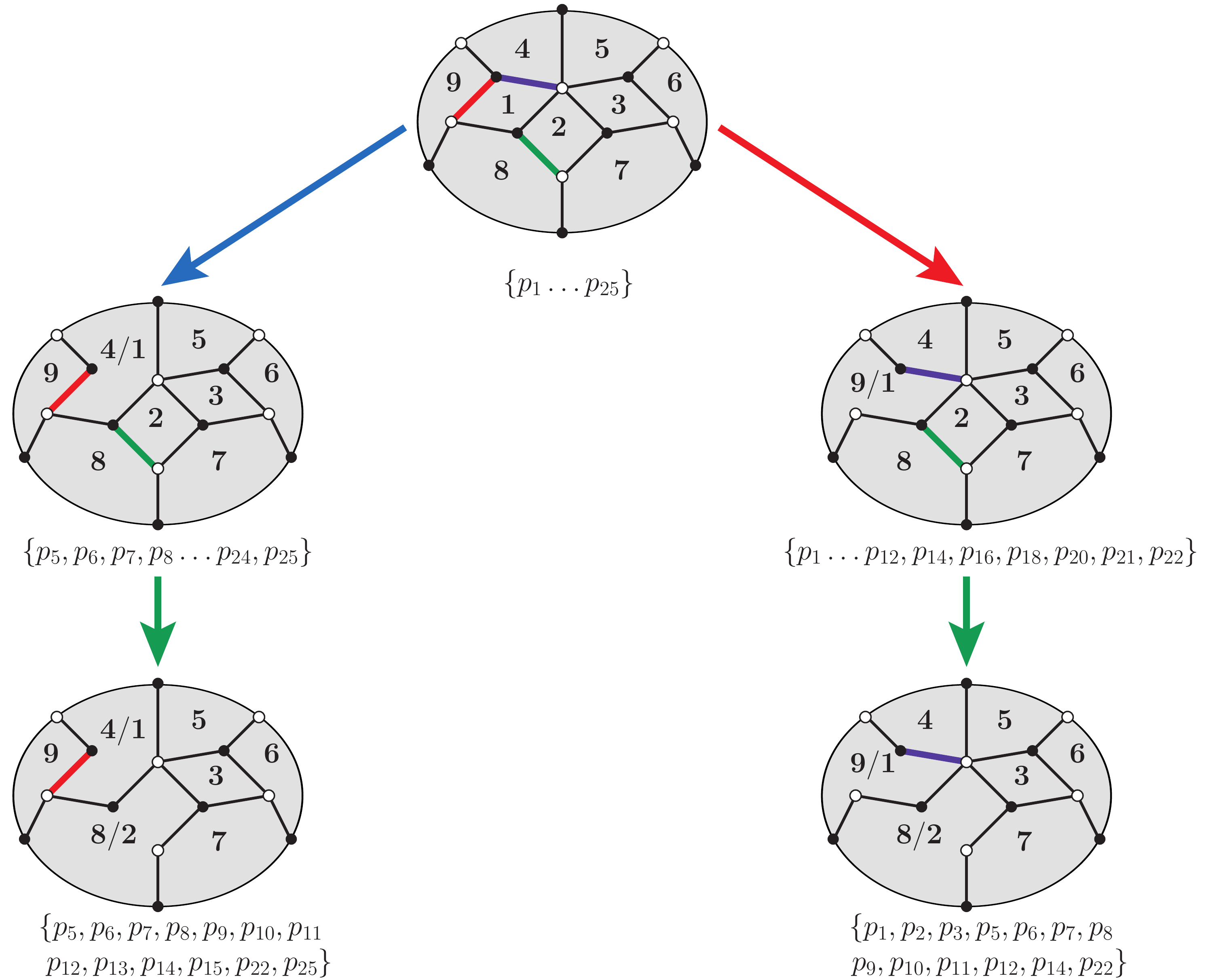}
\caption{The boundaries of the matching polytope are obtained by removing equivalence classes of edges. Two different 7-dimensional boundaries are shown, as well as two 6-dimensional sub-boundaries. There are many more boundaries of dimension 7 and 6, not shown in the figure.}
\label{fig:Gr26bound}
\end{center}
\end{figure}

Let us illustrate the above with a few explicit examples of boundaries of the $Gr_{2,6}$ example shown in \fref{dual_quiver}. The matrix $P$ for this example is given in \eqref{matching26}, and describes an 8-dimensional polytope. In \fref{fig:Gr26bound} we provide an example of 4 different subgraphs which are boundaries of the matching polytope, two of dimension 7 and two of dimension 6. The two boundaries of dimension 7 are obtained by removing the edges $X_{14}$ (shown in blue) and $X_{91}$ (shown in red), respectively. There are several other boundaries of dimension 7. The rows and columns erased in this process are highlighted in \eqref{matching26} in the corresponding colors, and under each graph in the figure we explicitly write which perfect matchings, i.e.\ which points in the matching polytope, are present in each of the boundaries. The two 6-dimensional faces we shall consider are obtained by further removing from the two previous boundaries the edge $X_{82}$, which is shown in green in \fref{fig:Gr26bound} and in \eqref{matching26}.

{\scriptsize
\be
\label{matching26}
P=
\left(
\begin{array}{  c | > {\columncolor{blue!40}} c > {\columncolor{blue!40}} c > {\columncolor{blue!40}} c > {\columncolor{blue!40}} c 
ccccc ccc > {\columncolor{red!40}}  c c > {\columncolor{red!40}} c  > {\columncolor{green!40}} c > {\columncolor{red!40}} c > {\columncolor{green!40}} c > {\columncolor{red!40}} c > {\columncolor{green!40}} c > {\columncolor{green!40}} cc > {\columncolor{red!40}} c > {\columncolor{red!40}} c > {\columncolor{red!40}} c}
 & p_{1} & p_{2} & p_{3} & p_{4} & p_{5} & p_{6} & p_{7} & p_{8} & p_{9} & p_{10} & p_{11} & p_{12} & p_{13} & p_{14} & p_{15} & p_{16} & p_{17} & p_{18} & p_{19} & p_{20} & p_{21} & p_{22} & p_{23} & p_{24} & p_{25} \\
 \hline
\rowcolor{blue!40}  X_{14} & 1 & 1 & 1 & 1 & 0 & 0 & 0 & 0 & 0 & 0 & 0 & 0 & 0 & 0 & 0 & 0 & 0 & 0 & 0 & 0 & 0 & 0 & 0 & 0 & 0 \\
X_{18} & 1 & 1 & 1 & 0 & 1 & 1 & 1 & 1 & 1 & 1 & 1 & 0 & 0 & 0 & 0 & 0 & 0 & 0 & 0 & 0 & 0 & 0 & 0 & 0 & 0 \\
X_{27} & 1 & 1 & 0 & 0 & 1 & 1 & 1 & 0 & 0 & 0 & 0 & 1 & 1 & 1 & 1 & 0 & 0 & 0 & 0 & 0 & 0 & 0 & 0 & 0 & 0 \\
X_{36} & 1 & 0 & 0 & 0 & 1 & 0 & 0 & 1 & 0 & 0 & 0 & 1 & 1 & 0 & 0 & 1 & 1 & 0 & 0 & 0 & 0 & 0 & 0 & 0 & 0 \\
X_{53} & 0 & 0 & 0 & 0 & 0 & 0 & 1 & 0 & 0 & 1 & 0 & 0 & 0 & 0 & 0 & 0 & 0 & 0 & 0 & 1 & 0 & 0 & 1 & 0 & 0 \\
X_{32} & 0 & 0 & 0 & 0 & 0 & 0 & 0 & 1 & 0 & 0 & 1 & 0 & 0 & 0 & 0 & 1 & 1 & 0 & 0 & 0 & 1 & 0 & 0 & 1 & 0 \\
X_{73} & 0 & 0 & 1 & 1 & 0 & 0 & 0 & 0 & 1 & 1 & 0 & 0 & 0 & 0 & 0 & 0 & 0 & 1 & 1 & 1 & 0 & 1 & 1 & 0 & 1 \\
X_{21} & 0 & 0 & 0 & 0 & 0 & 0 & 0 & 0 & 0 & 0 & 0 & 1 & 1 & 1 & 1 & 0 & 0 & 0 & 0 & 0 & 0 & 1 & 0 & 0 & 1 \\
 \rowcolor{green!40} X_{82} &0 & 0 & 0 & 1 & 0 & 0 & 0 & 0 & 0 & 0 & 0 & 0 & 0 & 0 & 0 & 1 & 1 & 1 & 1 & 1 & 1 & 0 & 1 & 1 & 0 \\
 \rowcolor{red!40} X_{91} &0 & 0 & 0 & 0 & 0 & 0 & 0 & 0 & 0 & 0 & 0 & 0 & 1 & 0 & 1 & 0 & 1 & 0 & 1 & 0 & 0 & 0 & 1 & 1 & 1 \\
X_{45} & 0 & 0 & 0 & 0 & 1 & 1 & 0 & 0 & 1 & 0 & 0 & 0 & 0 & 0 & 0 & 0 & 0 & 1 & 1 & 0 & 0 & 0 & 0 & 0 & 0 \\
 X_{49} &0 & 0 & 0 & 0 & 1 & 1 & 1 & 1 & 1 & 1 & 1 & 1 & 0 & 1 & 0 & 1 & 0 & 1 & 0 & 1 & 1 & 1 & 0 & 0 & 0 \\
 X_{67} &0 & 1 & 0 & 0 & 0 & 1 & 1 & 0 & 0 & 0 & 1 & 0 & 0 & 1 & 1 & 0 & 0 & 0 & 0 & 0 & 1 & 0 & 0 & 1 & 0 \\
X_{65} & 0 & 1 & 1 & 1 & 0 & 1 & 0 & 0 & 1 & 0 & 1 & 0 & 0 & 1 & 1 & 0 & 0 & 1 & 1 & 0 & 1 & 1 & 0 & 1 & 1 \\
X_{78} & 0 & 0 & 1 & 0 & 0 & 0 & 0 & 1 & 1 & 1 & 1 & 0 & 0 & 0 & 0 & 0 & 0 & 0 & 0 & 0 & 0 & 1 & 0 & 0 & 1 \\
X_{89} & 0 & 0 & 0 & 1 & 0 & 0 & 0 & 0 & 0 & 0 & 0 & 1 & 0 & 1 & 0 & 1 & 0 & 1 & 0 & 1 & 1 & 1 & 0 & 0 & 0
\end{array}
\right)
\ee
}

 %===============================================================================
%===============================================================================

\bigskip

%===============================================================================
\subsubsection{Identification of Perfect Matchings}
%===============================================================================

The final step required to obtain the combinatorial stratification of the Grassmannian is to \textit{identify} perfect matchings. This is motivated by the fact that, as can be seen in e.g.\ \sref{MatroidsandPolytopes}, multiple perfect matchings are associated with the same matroid base, and hence the same \pl coordinate of the Grassmannian. For this reason, we identify those perfect matchings which project to the same point in the matroid polytope. 

In general, this identification also identifies certain boundaries of the matching polytope. The identified boundaries may be of equal or different dimension; we refer to the former case as a \textit{horizontal} identification and to the latter as a \textit{vertical} identification. For vertical identifications, we always identify the boundaries involved with the lower-dimensional site of the poset. From a BFT perspective, this identification is a natural procedure: it is equivalent to identifying those graphs whose BFTs have the same moduli space.

In the example shown in \fref{fig:Gr26bound}, the identification is read off from \eref{eq:PerfOrientandMatroidPolytope}. The two 7-dimensional boundaries identify to 

\begin{eqnarray}
\begin{array}{c}
\{p_5,p_6,p_7,p_8,p_9,\boldsymbol{p_{10}},p_{11},p_{12},\boldsymbol{p_{13}},p_{14},\boldsymbol{p_{15}}, \\
\boldsymbol{p_{16}},\boldsymbol{p_{17}},p_{18},p_{19},\boldsymbol{p_{20}},\boldsymbol{p_{21}},p_{22},\boldsymbol{p_{23}},\boldsymbol{p_{24}},\boldsymbol{p_{25}}\}
\end{array} \quad \longrightarrow \quad \begin{array}{c}
\{p_1,p_2,p_3,p_5,p_6,p_7,p_8,p_9, \\
p_{11},p_{12},p_{14},p_{18},p_{19},p_{22}\}
\end{array} \quad \nonumber \\
\begin{array}{c}
\{p_1,p_2,p_3,p_4,p_5,p_6,p_7,p_8,p_9,\boldsymbol{p_{10}}, \\
p_{11},p_{12},p_{14},\boldsymbol{p_{16}},p_{18},\boldsymbol{p_{20}},\boldsymbol{p_{21}},p_{22}\}
\end{array} \quad \longrightarrow \quad \begin{array}{c}
\{p_1,p_2,p_3,p_4,p_5,p_6,p_7,p_8, \\
p_9,p_{11},p_{12},p_{14},p_{18},p_{22}\}
\end{array} \quad 
\end{eqnarray}
where we highlighted in bold those perfect matchings that get identified to a different perfect matching. The two 6-dimensional boundaries identify to

\begin{eqnarray}
\begin{array}{c}
\{p_5,p_6,p_7,p_8,p_9,\boldsymbol{p_{10}},p_{11}, \\
p_{12},\boldsymbol{p_{13}},p_{14},\boldsymbol{p_{15}},p_{22},\boldsymbol{p_{25}}\}
\end{array} \quad \longrightarrow \quad \begin{array}{c}
\{p_1,p_2,p_3,p_5,p_6,p_7,p_8, \\
p_9,p_{11},p_{12},p_{14},p_{22}\}
\end{array} \quad \phantom{.}\nonumber \\
\begin{array}{c}
\{p_1,p_2,p_3,p_5,p_6,p_7,p_8, \\
p_9,\boldsymbol{p_{10}},p_{11},p_{12},p_{14},p_{22}\}
\end{array} \quad \longrightarrow \quad \begin{array}{c}
\{p_1,p_2,p_3,p_5,p_6,p_7,p_8, \\
p_9,p_{11},p_{12},p_{14},p_{22}\}
\end{array} \quad .
\end{eqnarray}
Here we see an explicit example of a horizontal identification: the two 6-dimensional boundaries get identified.

Performing this identification on the entire matching polytope poset, we end up with the combinatorial stratification. We point out that at the last level, of dimension zero, the number of boundaries becomes the number of \pl coordinates, as required by the very nature of the stratification of the Grassmannian. For several examples where the stratification has been done in full, for both planar and non-planar cases, see \cite{Franco:2013nwa}.

\bigskip

%===============================================================================
\section{Cluster Transformations from QFT}
%===============================================================================

Interestingly, the BFT interpretation of bipartite graphs permits the derivation of cluster transformations of face weights under square moves from a fundamental property of Seiberg dual theories: the matching of their moduli spaces \cite{Franco:2013pg}.\footnote{Strictly speaking, Seiberg duality holds only for non-Abelian gauge theories. Here we focus on classical, Abelian BFTs and study the associated modification of the graph. Since Seiberg duality corresponds to a local transformation of the graph, our discussion can be promoted to non-Abelian theories.} Cluster algebras were originally introduced in \cite{MR1887642,MR2295199}.

%===============================================================================
\begin{figure}[h]
\begin{center}
\includegraphics[width=13cm]{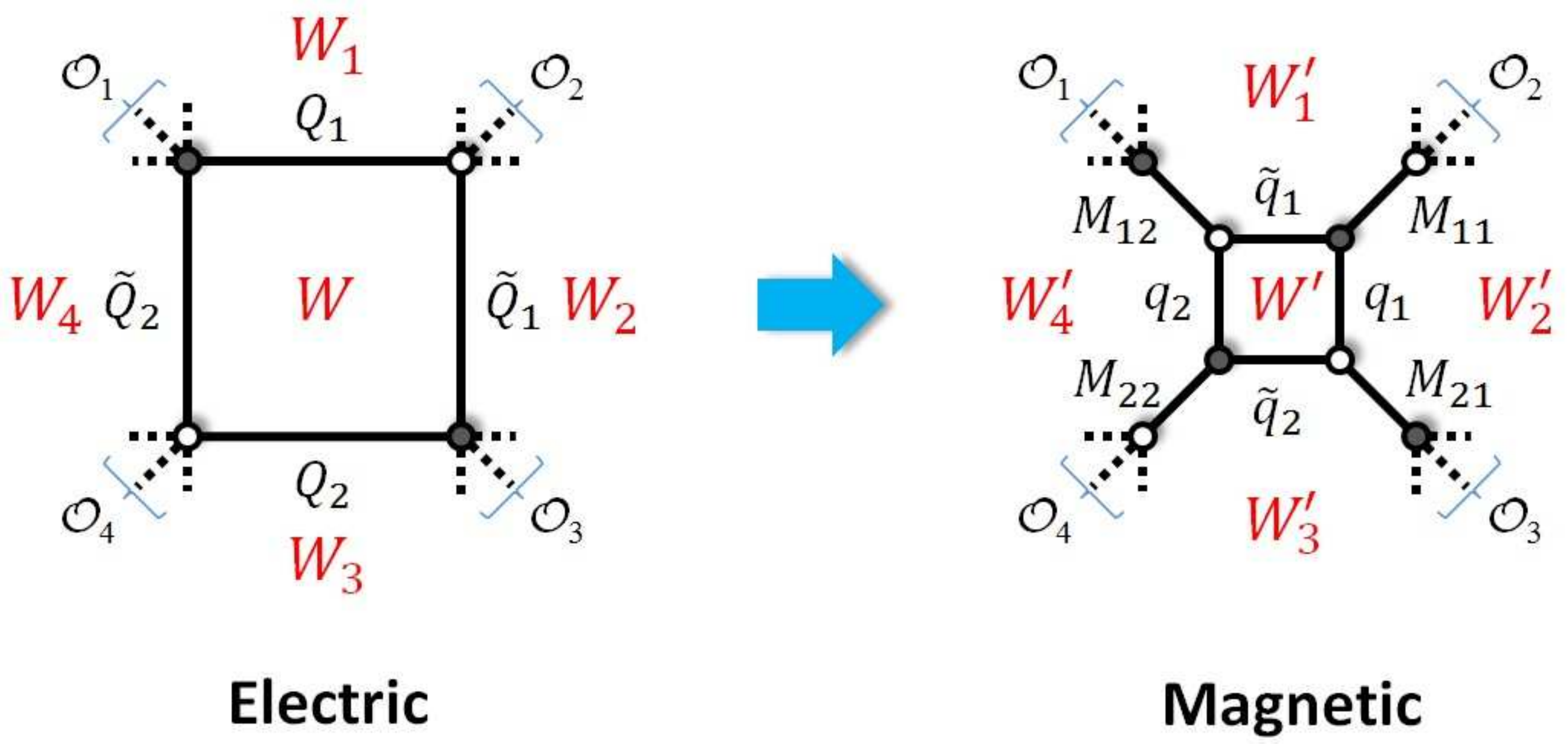}
\caption{The graphs for the electric and magnetic theories are related by a square move.}
\label{cluster}
\end{center}
\end{figure}
%===============================================================================

It is sufficient to focus on the {\it electric} configuration shown in \fref{cluster}, which represents a piece of a general BFT. It corresponds to standard supersymmetric (SQCD) with $N_f=2N_c$, extended by additional operators $\mathcal{O}_\alpha$, $\alpha=1,\ldots,4$. The $\mathcal{O}_\alpha$ operators can represent single chiral superfields or products of them, i.e.\ single or multiple edges. In addition, the theory has the following superpotential

\beq
W_{el}=-\tilde{Q}_2 Q_1  \mathcal{O}_1 +\tilde{Q}_1 Q_1  \mathcal{O}_2 -\tilde{Q}_1 Q_2  \mathcal{O}_3 +\tilde{Q}_2 Q_2 \mathcal{O}_4.
\label{W_el}
\eeq

\fref{cluster} also shows the {\it magnetic} theory obtained by Seiberg dualizing the gauge group associated to the square face. Following the discussion in \sref{section_graphical_dynamics}, this duality is graphically implemented in terms of a square move. The resulting theory is the usual magnetic dual of $N_f=2N_c$ SQCD, with operators $\mathcal{O}_\alpha$ and superpotential 

\beq
W_{mag}=[-M_{12}  \mathcal{O}_1 +M_{11}  \mathcal{O}_2 -M_{21}  \mathcal{O}_3 +M_{22} \mathcal{O}_4]+ \tilde{q}_1 q_2  M_{12} - \tilde{q}_1 q_1  M_{11}+ \tilde{q}_2 q_2  M_{22}- \tilde{q}_2 q_1  M_{21}.
\label{W_mag}
\eeq

Let us now consider the moduli spaces of the electric and magnetic theory. When doing so, we do not impose vanishing of F-terms for the $\mathcal{O}_\alpha$'s. When they represent external legs, this is the special treatment discussed in \sref{section_master_and_moduli_spaces}, otherwise, this amounts to focusing on an appropriate patch of the moduli space. Following our general discussion, perfect matchings provide nice smooth parametrizations of these moduli spaces. The perfect matchings for both theories, and indeed the toric diagrams for the moduli spaces, are neatly encoded in the following characteristic polynomials

\beq
\begin{array}{ccl}
\mathcal{P}_e  & = & (Q_1 Q_2 + \tilde{Q}_1 \tilde{Q}_2) + \mathcal{O}_1 \mathcal{O}_2 Q_2 + \mathcal{O}_1 \mathcal{O}_4 \tilde{Q}_1 + \mathcal{O}_2 \mathcal{O}_3 \tilde{Q}_2 + \mathcal{O}_3 \mathcal{O}_4 Q_1+ \mathcal{O}_1 \mathcal{O}_2 \mathcal{O}_3 \mathcal{O}_4 ,
\\ \\
\mathcal{P}_m & = & M_{11}M_{12}M_{21}M_{22}+\mathcal{O}_1 \mathcal{O}_2 M_{21} M_{22} \tilde{q}_1 + \mathcal{O}_1 \mathcal{O}_4 M_{11} M_{21} q_2 + \mathcal{O}_2 \mathcal{O}_3 M_{12} M_{22} q_1 \\
& + & \mathcal{O}_3 \mathcal{O}_4 M_{11} M_{12} \tilde{q}_2 +  \mathcal{O}_1 \mathcal{O}_2 \mathcal{O}_3 \mathcal{O}_4 (q_1 q_2 + \tilde{q}_1 \tilde{q}_2),
\end{array}
\label{characteristic_polynomials}
\eeq
where every term corresponds to a perfect matching.

The moduli space of a gauge theory is invariant under Seiberg duality. Matching the moduli spaces of the electric and magnetic theories implies equating the terms in both polynomials with the same $\mathcal{O}_\alpha$ content, since the $\mathcal{O}_\alpha$'s are insensitive to the duality. Doing so leads to {\it formal} relations between the fields, equivalently edge weights, in the two theories.\footnote{These relations should be understood as leading to relations between perfect matchings which, in turn, guarantee the matching of all gauge invariant operators in the chiral ring of the two theories. They should not be thought of as operator relations at the level of quiver fields.}

Let us investigate the effect of these relations on oriented paths. Oriented paths on the bipartite graph are related to oriented edge weights, i.e.\ with an orientation going from white to black nodes, as follows

\beq
v(\gamma)=\prod_{i=1}^{k-1} {X(\ww_i,\bb_i)\over X(\ww_{i+1},\bb_i)} \, ,
\label{flux_v}
\eeq
where the product runs over the path $\gamma$ and $\bb_i$ and $\ww_j$ denote black and white nodes. Subindices indicate the nodes connected by the corresponding edge when moving along $\gamma$.

Combining \eref{flux_v} with the mapping of edge weights derived from matching the characteristic polynomials, we obtain the following relations for the clockwise oriented face weights shown in \fref{cluster}:
\beq
\begin{array}{cclcccl}
W_1' & = & W_1 (1+W) & \ \ \ \ \ \ & W_2' & = & W_2 (1+W^{-1})^{-1} \\ 
W_3' & = & W_3 (1+W) & \ \ \ \ \ \ & W_4' & = & W_4 (1+W^{-1})^{-1} .
\end{array}
\eeq
These are the standard cluster transformations of face weights \cite{2006math09764P}. Remarkably, we have derived them from the invariance of the moduli space under Seiberg duality.

\bigskip

%===============================================================================
\section{String Theory Embedding}
%===============================================================================

\label{section_string_embedding}

In this section we will outline the string theory embedding of a sub-class of BFTs. This class corresponds to BFTs associated to graphs that can be drawn on a plane.\footnote{Notice that this class is larger than the one for {\it planar} graphs. For us, planar graphs are those that live on a disk, i.e.\ they have a single boundary, which implies that all external nodes can be sent to infinity without introducing crossings between edges.}

There are various motivations for searching for a string embedding of BFTs. By doing so, one often gets a better understanding of dualities and computational control over strongly coupled regimes via the gauge/gravity correspondence. Such embeddings might also hint to possible applications of the quantum field theories to elucidate the poorly understood 6d theories on M5-brane, e.g.\ via deconstruction \cite{ArkaniHamed:2001ie}. Finally, additional consistency constraints that might arise in the string theory context might also narrow the space of theories which are ``better behaved". For the specific case of BFTs, one also gets additional physical motivations for the special treatment of the chiral fields associated to external legs.

\bigskip

%===============================================================================
\subsection{General Strategy}
%===============================================================================

\begin{figure}[h]
\begin{center}
\includegraphics[width=4.6cm]{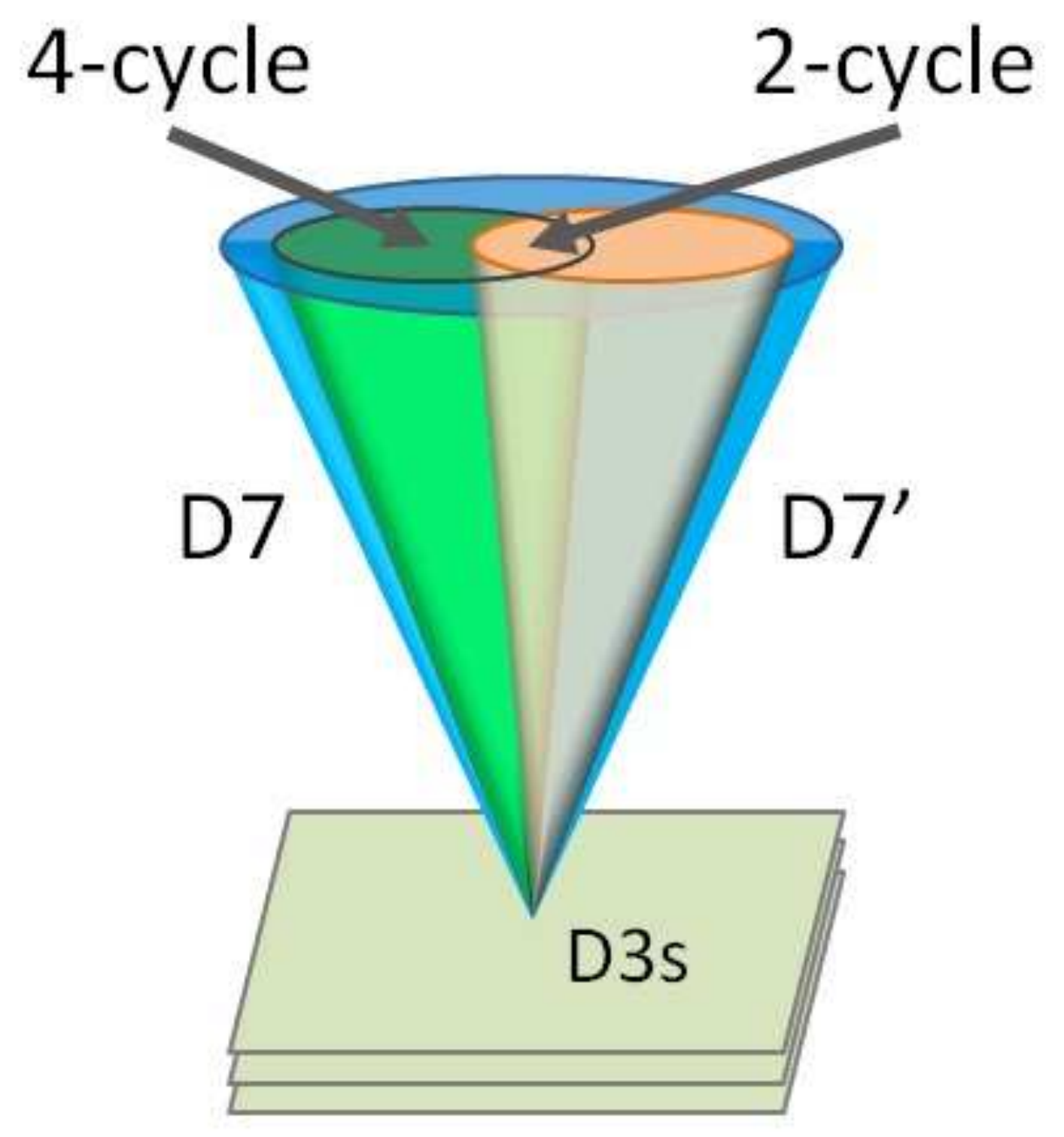}
\caption{Schematic configuration of D3-branes and flavor D7-branes on a CY 3-fold.}
\label{basic_configuration}
\end{center}
\end{figure}

In \cite{Franco:2013ana}, BFTs were engineered in terms of fractional D3-branes and flavor D7-branes at non-compact, toric, singular CY 3-folds.\footnote{A string theory embedding for the theories in \cite{Xie:2012mr} was given in \cite{Heckman:2012jh}.} Both types of D-branes extend along the four dimensions on which the BFT lives. Flavor branes, in addition, wrap non-compact 4-cycles in the CY. The basic ingredients in such a configuration are illustrated in \fref{basic_configuration}. A useful device for describing D-branes in toric CY 3-folds is the underlying dimer model \cite{Hanany:2005ve,Franco:2005rj,Franco:2005sm,Kennaway:2007tq}.

%===============================================================================
%===============================================================================

What is the main challenge one faces when trying to find a string theory realization of BFTs? It is necessary to develop a detailed understanding of the gauge theories on configurations of D-branes at singularities in order to determine whether a spectrum and superpotential interactions consistent with a bipartite graph can arise from these systems. In order to do so, a comprehensive framework for determining the gauge theories on general D-brane configurations on toric singularities was introduced in \cite{Franco:2013ana}, extending previous work in \cite{Franco:2006es}. This technology has a wide range of applications, such as local approaches to string phenomenology, which significantly exceeds its uses in the context of BFTs.

\begin{figure}[h]
\begin{center}
\includegraphics[width=8cm]{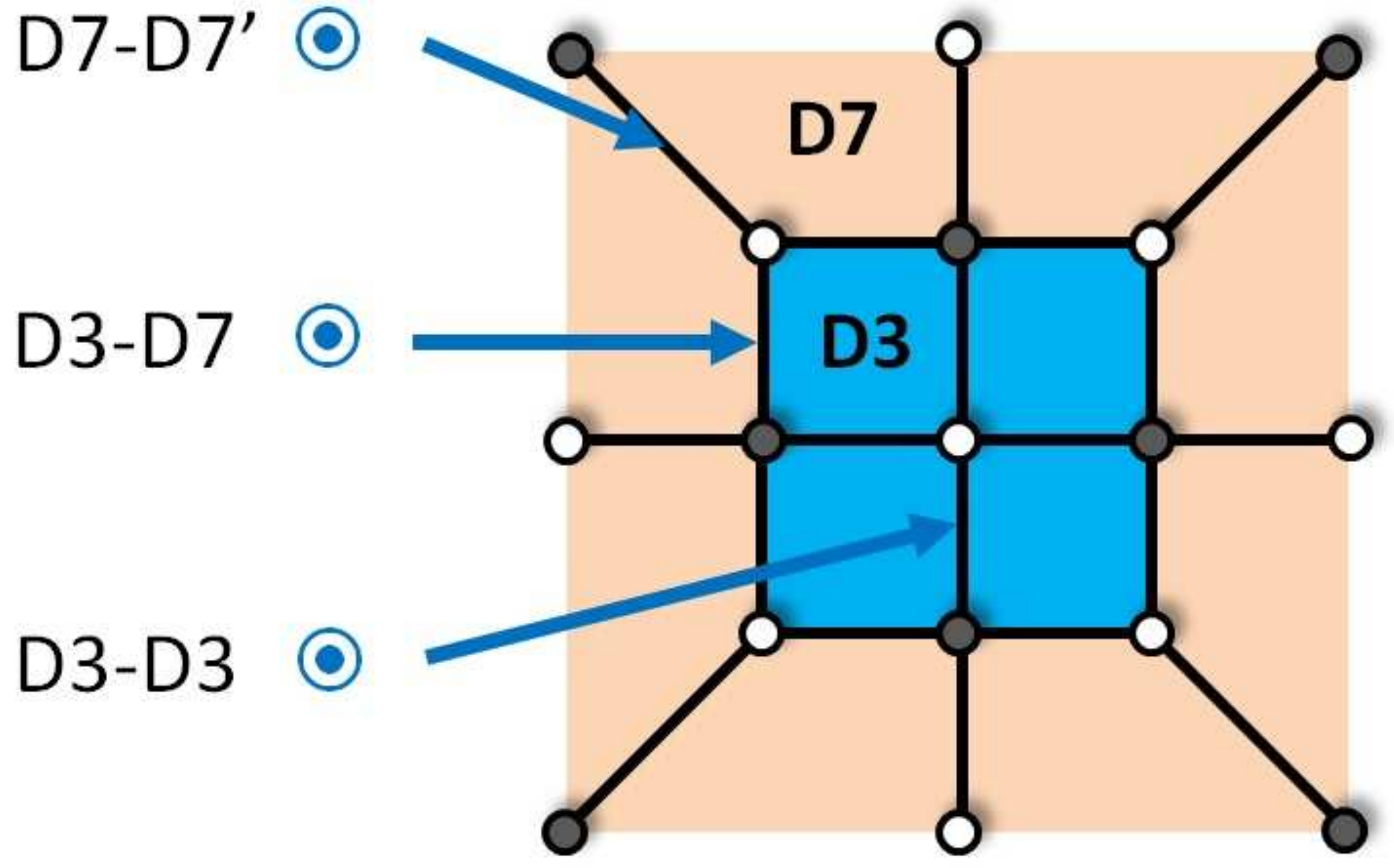}
\caption{An example of a BFT indicating the types of D-branes and open string sectors giving rise to its faces and edges.}
\label{D-brane_sectors}
\end{center}
\end{figure}

It is useful to distinguish edges according to the types of faces, internal or external, they separate. \fref{D-brane_sectors} schematically illustrates the D-brane origin of some of the basic features of BFTs. 

%===============================================================================
%===============================================================================

In more detail, we have:

\bigskip

\begin{itemize}
\item {\bf Internal faces:} they arise from fractional D3-branes and correspond to occupied faces in an underlying dimer model. 

\item {\bf External faces:} they correspond to stacks of non-compact D7-branes. The 8d gauge symmetry on the worldvolume of these D7-branes becomes a global symmetry from the viewpoint of the four dimensions in which the BFT lives, due to non-compact extension of the D7-branes along the internal dimensions.

\item {\bf Internal/internal edges:} this part of the bipartite graph is directly inherited from the underlying dimer model. These edges correspond to D3-D3 states. 

\item {\bf External/external edges:} we have been referring to these edges as external legs. The associated fields transform in bifundamental representations of global symmetry groups. They correspond to D7-D7' states. These fields live at the 6d intersections between D7-branes wrapping different 4-cycles, as shown in \fref{basic_configuration}. They have a non-compact support and hence are non-dynamical fields from a 4d viewpoint providing further motivation for how we treat them. They can also arise between D7-branes which wrap the same non-compact 4-cycle, and thus have an 8d support.

\item {\bf Internal/external edges:} the corresponding fields are typically called {\it flavors}, since they transform in the fundamental or antifundamental representation of a gauge group. In addition, they transform in antifundamental or fundamental representation of a global symmetry group, respectively. They arise from states in D3-D7 sectors. 
\end{itemize}

\bigskip

%===============================================================================
\subsection{The Mirror}
%===============================================================================

A powerful tool for investigating D-branes over toric singularities is their {\it mirror}. For the theories at hand, the mirror is given by a $\Sigma_w \times \mathbb{C}^*$ fibration over the $w$ complex plane:
\begin{eqnarray}
P(x,y) & = & w  \nonumber \\
u \, v & = & w \, .
\end{eqnarray}
For every point $w$, there is a Riemann surface $\Sigma_w$ corresponding to $P(x,y) = w$. $P(x,y)=\sum a_{n_1,n_2} x^{n_1} y^{n_2}$ is the characteristic polynomial associated to the toric CY under consideration, i.e.\ it contains a term for every point in the toric diagram.

The most important aspects of D-branes in this geometry are captured by the Riemann surface $\Sigma$ at $w=0$. D3-branes in the original CY are mapped to D6-branes over compact 3-cycles in the mirror, which are projected down to compact 1-cycles on $\Sigma$. Similarly, flavor D7-branes are turned into D6-branes over non-compact 3-cycles, which descend to non-compact 1-cycles on $\Sigma$.

The brane tiling associated to the original CY, a bipartite graph $G$ on $T^2$, is turned into a new bipartite graph $\tilde{G}$ on $\Sigma$ by the {\it untwisting map}. We refer the reader to \cite{Feng:2005gw} for details of this correspondence. The untwisting map turns faces and zig-zag paths of $G$ into zig-zag paths and faces of $\tilde{G}$, respectively. Faces in $\tilde{G}$  surround the punctures in $\Sigma$. Gauge groups in the resulting quiver, i.e.\ fractional D3-branes in the original dimer, correspond to zig-zag paths on $\tilde{G}$.

\bigskip

%===============================================================================
\paragraph{Short Embeddings.}
%===============================================================================

Without going into technical details, let us discuss some basic properties of flavor D7-branes. As we explained, they correspond to non-compact 1-cycles which extend to infinity through a pair of punctures in $\Sigma$. The intersections between these 1-cycles and $\tilde{G}$ determine the field content and interactions of the gauge theory. 

%===============================================================================
\begin{figure}[h]
\begin{center}
\includegraphics[width=14cm]{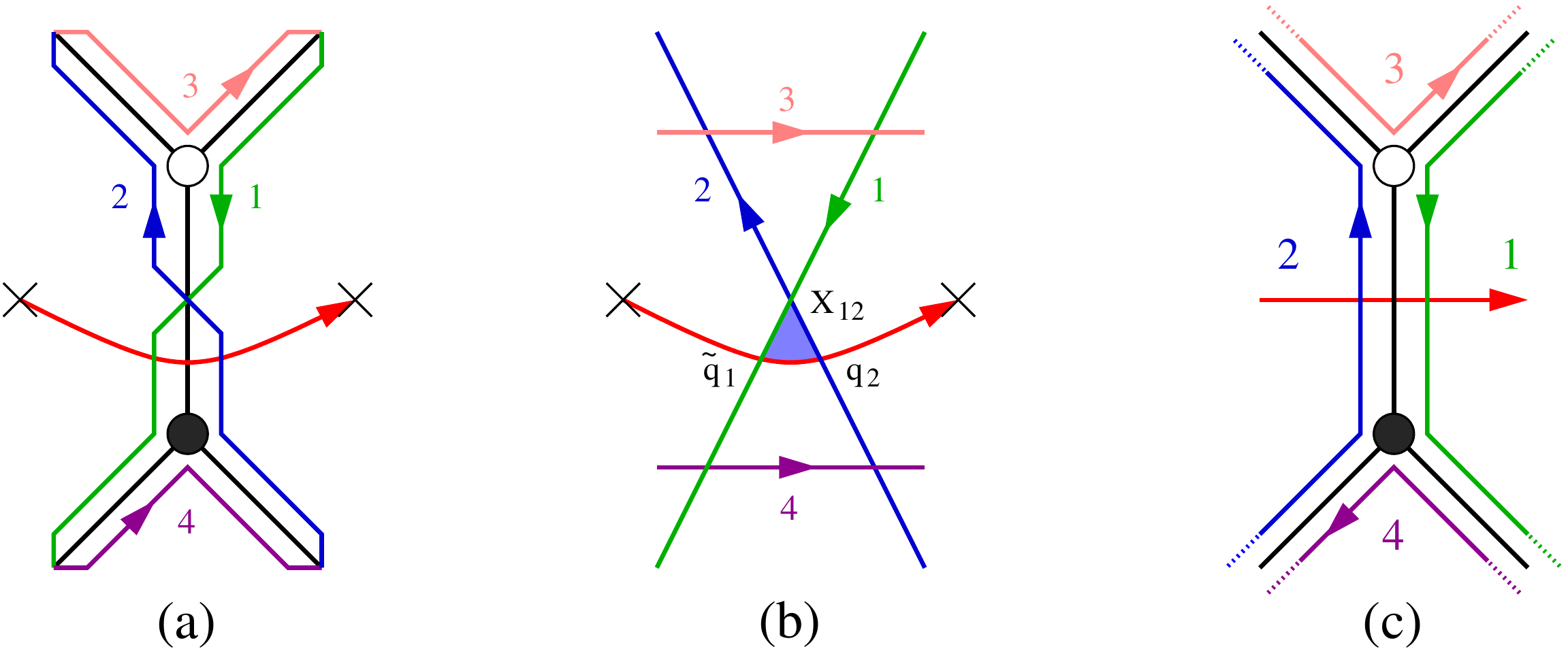}
\caption{Basic configuration for a short embedding D7-brane. a) Mirror picture showing it as an oriented cycle between two punctures. b) The same configuration eliminating the underlying graph $\tilde{G}$. In blue, we show the disk supporting the instanton giving rise to a superpotential term $W_{3\,7}=\tilde{q}_1X_{12}q_2$. c) The corresponding portion of the original dimer, obtained by untwisting the mirror.}
\label{mirror_short_embedding}
\end{center}
\end{figure}
%===============================================================================

We refer to the simplest type of flavor D7-branes as {\it short embeddings}. A short embedding corresponds to a non-compact 1-cycle crossing a single edge in $\tilde{G}$. Equivalently, such a D7-brane crosses two zig-zag paths in $\tilde{G}$, i.e.\ two D3-brane faces in the original dimer model. These intersections give rise to two flavors, i.e.\ chiral multiplets in the fundamental and antifundamental representations of the two D3-brane gauge factors. Each D7-brane 1-cycle has an orientation, as shown in \fref{mirror_short_embedding}.b, which we pick such that in combination with the two zig-zag paths it forms the oriented boundary of disk supporting a worldsheet instanton that generates a superpotential coupling between the flavors and D3-D3 fields of the form

\beq
W_{3\,7}= {\tilde q}_{73} X_{33'} q_{3'7} .
\label{supo0}
\eeq
Short embedding D7-branes can be represented in the original dimer by an oriented arrow across an edge going in the opposite orientation to the corresponding bifundamental. 

The main ingredients of a short embedding are presented in \fref{mirror_short_embedding}. Zig-zag paths are shown in the double line notation of \cite{Feng:2005gw} and punctures are indicated with crosses.

\begin{figure}[h]
\begin{center}
\includegraphics[width=12cm]{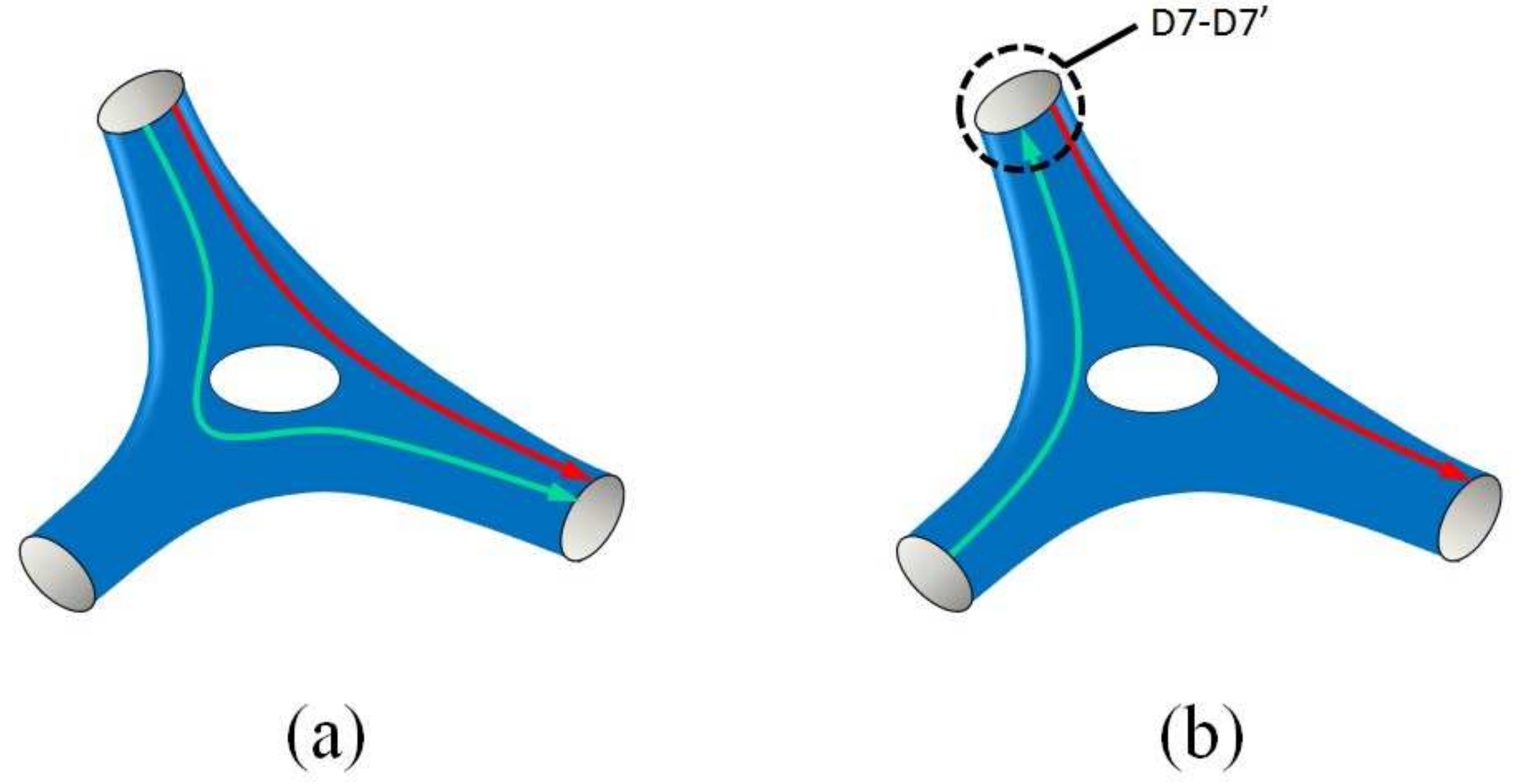}
\caption{Flavor D7-branes map to non-compact 1-cycles on $\Sigma$ going to infinity through two punctures. a) Branes with the same pair of punctures but different trajectories in the bulk differ in the worldvolume gauge field. b) D7-D7' states arise at `intersections' between D7-branes sharing punctures.}
\label{D7s_punctures_and_intersections}
\end{center}
\end{figure}

As shown in \fref{D7s_punctures_and_intersections}.a, it is possible for two 1-cycles to share the same pair of punctures but differ in their trajectories along the bulk of $\Sigma$ or, more precisely, the edges in $\tilde{G}$ they cross. Such trajectories correspond to D7-branes wrapped over the same geometric 4-cycle but that differ in their worldvolume gauge field. They generalize the orbifold case of D7-branes wrapping the same 4-cycles but differing in their Chan-Paton factors.

%===============================================================================
%===============================================================================

The second general observation we would like to make regards the mirror description of D7-D7' states. These states arise at intersections between D7-branes, more precisely they stretch between pairs of D7-branes with opposite orientations `intersecting' at a single puncture in $\Sigma$, as shown in an example in \fref{D7s_punctures_and_intersections}.b. Given a pair of D7-branes $A$ and $B$, the D7-D7' field $Y_{AB}$ has the following superpotential coupling to flavors

\beq
W'_{\rm 3\, 7} = q_{iA}\, Y_{AB}\, \tilde{q}_{Bi}.
\label{W_73-33-37}
\eeq
D7-D7' states can also arise from the 8d worldvolume of D7-branes wrapped over the same 4-cycle but with different gauge bundles, a situation we described above. In this case, the corresponding 1-cycles share two punctures but differ in the bulk.

\bigskip

%===============================================================================
\paragraph{Long Embeddings.}
%===============================================================================

It is possible to understand more general D7-branes associated to {\it long embeddings}. Generalizing embeddings associated to a single bifundamental field, long embeddings  correspond to paths of the form $\mathcal{O}_{i_0 i_n}=X_{i_0 i_1} X_{i_1 i_2} \ldots X_{i_{n-1}i_n}$, where consecutive fields share both a common gauge group and a puncture in $\Sigma$.\footnote{Open paths in which consecutive fields do not share a puncture can be deformed into paths in which they do so by using F-term relations.} First, we should consider the multiple short embedding D7-branes associated to each $X_{i_{\mu-1} i_\mu}$. The gauge theory of long-embedding strings is straightforwardly obtained by giving non-vanishing vacuum expectation values to the D7-D7' fields existing at each common puncture between consecutive branes and taking the low energy limit. The resulting D7-branes are represented by combining the arrows representing the original short embeddings.

\bigskip

In order for brane configurations to be consistent, they must satisfy the cancellation of RR tadpoles. We refer the reader to \cite{Franco:2013ana} for a detailed discussion of such constraints for the theories considered in this section.

\bigskip

%===============================================================================
\subsection{Explicit Examples}
%===============================================================================

\begin{figure}[h]
 \centering
 \begin{tabular}[c]{ccc}
 \epsfig{file=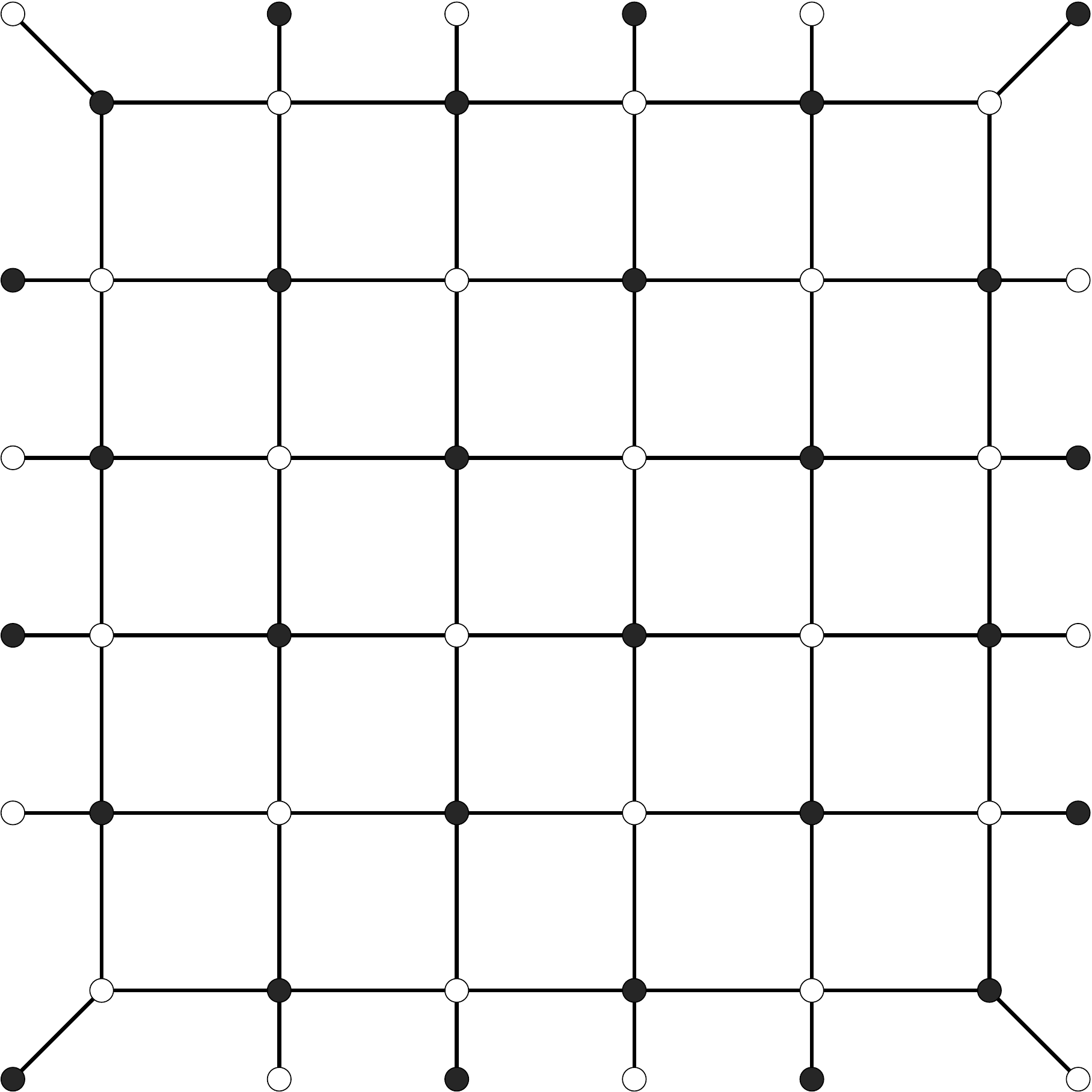,width=0.4\linewidth,clip=} & \ \ \ \ \ &
\epsfig{file=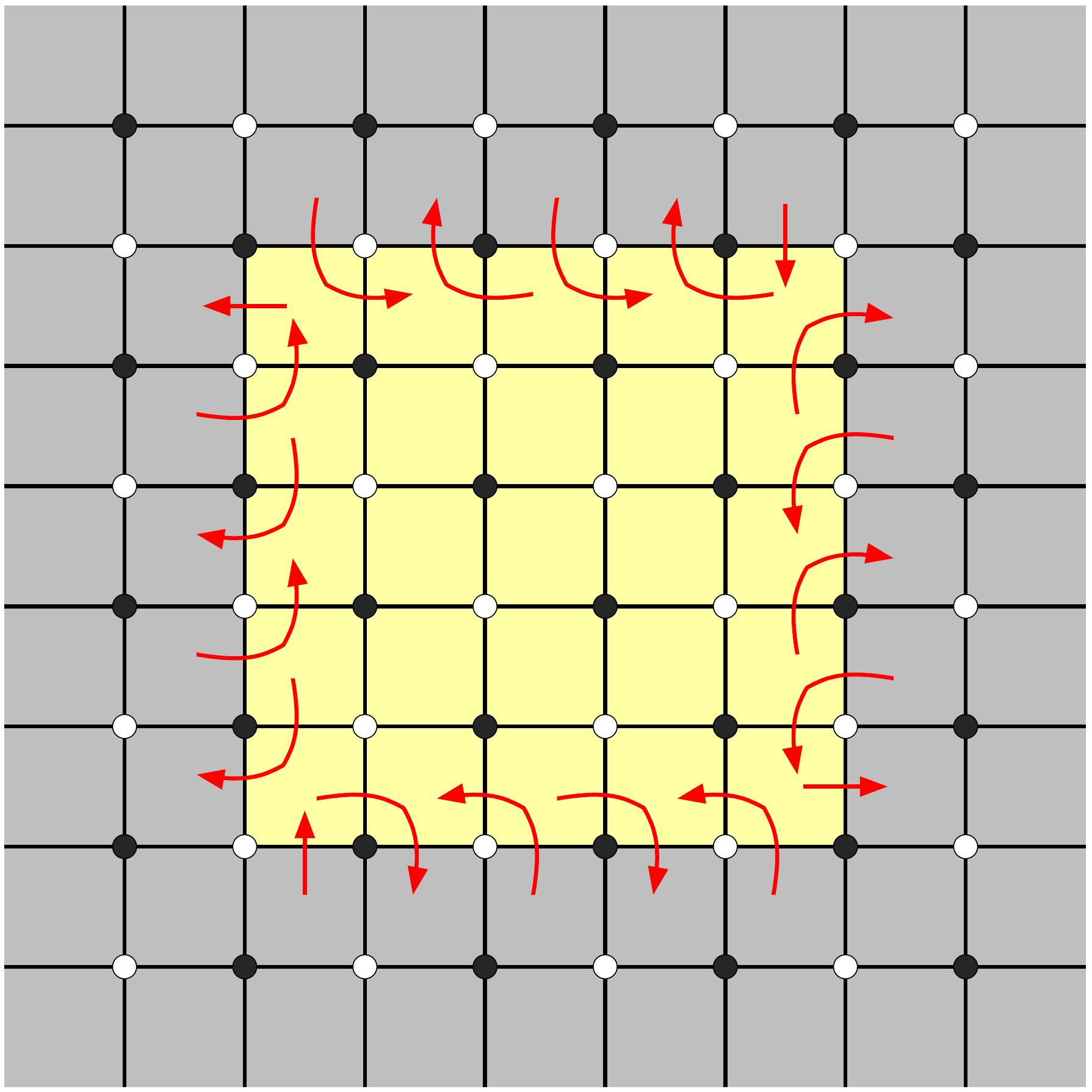,width=0.4\linewidth,clip=} \\ 
(a) & & (b)
 \end{tabular}
\caption{a)	A BFT in an infinite family of models corresponding to taking an arbitrary number of squares in the horizontal and vertical directions. b) Brane configuration engineering this gauge theory. Occupied and empty faces in the underlying dimer model are shown in yellow and grey, respectively. Arrows indicate D7-branes.}
\label{dimer_classB_higgsed} 
\end{figure}

The ideas reviewed in the previous section were exploited in \cite{Franco:2013ana} to construct infinite families of BFTs. \fref{dimer_classB_higgsed}, taken from \cite{Franco:2013ana}, shows an example of an infinite class of BFTs and the corresponding D-brane configuration. These theories can have an arbitrary number of squares in the horizontal and vertical directions. They can be engineered with D-branes on appropriately large $\mathbb{Z}_N \times \mathbb{Z}_M$ orbifolds of the conifold, which correspond to an underlying dimer model with a unit cell containing $N\times M$ squares. 

\bigskip

%===============================================================================
 %===============================================================================

\bigskip

%===============================================================================
\section*{Acknowledgements}
%===============================================================================

We would like to thank N. Arkani-Hamed, J. Bourjaily and J. Trnka  for useful discussions. We are also extremely grateful to R.-K. Seong and A. Uranga for enjoyable collaborations on some of the work presented in this review. We also thank the editors of the special issue of Journal of Physics A on
``Cluster Algebras in Mathematical Physics" for inviting us to contribute to it. The work of S. F. and D. G. is supported by the U.K. Science and Technology Facilities Council (STFC). A.M. acknowledges funding by the Durham International Junior Research Fellowship. S. F. would like to thank the generous hospitality of the Institute for Advanced Study, the Kavli Institute for Theoretical Physics and the Stanford Institute for Theoretical Physics during the preparation of this work. The research of S. F. was also supported in part by the National Science Foundation under Grant No. NSF PHY11-25915.

\bigskip

\appendix

%===============================================================================
\section{Perfect Matchings for the $Gr_{2,6}$ Example}
%===============================================================================

\label{app:PMlist}

For reference, here we list the perfect matchings for the bipartite graph in \fref{dual_quiver}. We have picked a notation in which subindices indicate the pair of faces separated by an edge.

\bea
&
p_1=\{ \alpha_{14}, \alpha_{18}, \alpha_{27},\alpha_{36} \} 
\qquad \qquad \qquad 
& p_2=\{ \alpha_{14}, \alpha_{18}, \alpha_{27},\alpha_{65}, \alpha_{67} \} 
\nonumber \\
&
p_3=\{ \alpha_{14}, \alpha_{18}, \alpha_{65},\alpha_{73}, \alpha_{78} \} 
\qquad \qquad 
& p_4=\{ \alpha_{14}, \alpha_{65}, \alpha_{73},\alpha_{82}, \alpha_{89} \} \nonumber  \\
&
p_5=\{ \alpha_{18}, \alpha_{27}, \alpha_{36},\alpha_{45}, \alpha_{49} \} 
\qquad \qquad 
& p_6=\{ \alpha_{18}, \alpha_{27}, \alpha_{45},\alpha_{49}, \alpha_{65} , \alpha_{67} \} \nonumber  \\
&
p_7=\{ \alpha_{18}, \alpha_{27}, \alpha_{49},\alpha_{53}, \alpha_{67} \} 
\qquad \qquad 
& p_8=\{ \alpha_{18}, \alpha_{32}, \alpha_{36},\alpha_{49},  \alpha_{78} \} \nonumber  \\
&
p_9=\{ \alpha_{18}, \alpha_{45}, \alpha_{49},\alpha_{65}, \alpha_{73} , \alpha_{78} \} 
\qquad 
& p_{10}=\{ \alpha_{18}, \alpha_{49}, \alpha_{53},\alpha_{73},  \alpha_{78} \} \nonumber  \\
&
p_{11}=\{ \alpha_{18}, \alpha_{32}, \alpha_{49},\alpha_{65}, \alpha_{67} , \alpha_{78} \} 
\qquad 
& p_{12}=\{ \alpha_{21}, \alpha_{27}, \alpha_{36},\alpha_{49},  \alpha_{89} \} \nonumber  \\
&
p_{13}=\{ \alpha_{21}, \alpha_{27}, \alpha_{36},\alpha_{91} \} 
\qquad \qquad \qquad 
& p_{14}=\{ \alpha_{21}, \alpha_{27}, \alpha_{49},\alpha_{65},  \alpha_{67} ,\alpha_{89} \} \nonumber  \\
&
p_{15}=\{ \alpha_{21}, \alpha_{27}, \alpha_{65}, \alpha_{67} , \alpha_{91} \} 
\qquad \qquad 
& p_{16}=\{ \alpha_{32}, \alpha_{36}, \alpha_{49},\alpha_{82},  \alpha_{89} \} \nonumber  \\
&
p_{17}=\{ \alpha_{32}, \alpha_{36}, \alpha_{82}, \alpha_{91} \} 
\qquad \qquad \qquad 
& p_{18}=\{ \alpha_{45}, \alpha_{49}, \alpha_{65},\alpha_{73},  \alpha_{82}, \alpha_{89} \} \nonumber  \\
&
p_{19}=\{ \alpha_{45}, \alpha_{65}, \alpha_{73}, \alpha_{82} , \alpha_{91} \} 
\qquad \qquad 
& p_{20}=\{  \alpha_{49}, \alpha_{53},\alpha_{73},  \alpha_{82}, \alpha_{89} \} \nonumber  \\
&
p_{21}=\{ \alpha_{32}, \alpha_{49}, \alpha_{65}, \alpha_{67},\alpha_{82},  \alpha_{89} \}
\qquad 
&
p_{22}=\{ \alpha_{21}, \alpha_{49}, \alpha_{65}, \alpha_{73},\alpha_{78},  \alpha_{89} \} \nonumber \\
&
p_{23}=\{ \alpha_{53}, \alpha_{73}, \alpha_{82}, \alpha_{91} \}
\qquad \qquad \qquad 
&
p_{24}=\{ \alpha_{32}, \alpha_{65}, \alpha_{67}, \alpha_{82},\alpha_{91} \}\nonumber \\
&
p_{25}=\{ \alpha_{21}, \alpha_{65}, \alpha_{73},\alpha_{78},  \alpha_{91} \} 
\qquad \qquad &
\eea

\bigskip

\bibliographystyle{JHEP}
\bibliography{matroref}
\end{document}

%% file: pref.tex
\newcommand{\be}{\begin{equation}}
\newcommand{\ee}{\end{equation}}
\newcommand{\beq}{\begin{equation}}
\newcommand{\beql}[1]{\begin{equation}\label{#1}}
\newcommand{\eeq}{\end{equation}}
\newcommand{\ba}{\begin{array}}
\newcommand{\ea}{\end{array}}
\newcommand{\bea}{\begin{eqnarray}}
\newcommand{\beal}[1]{\begin{eqnarray}\label{#1}}
\newcommand{\eea}{\end{eqnarray}}
\newcommand{\ben}{\begin{enumerate}}
\newcommand{\een}{\end{enumerate}}
\newcommand{\bean}{\begin{eqnarray*}}
\newcommand{\eean}{\end{eqnarray*}}
\newcommand{\eref}[1]{(\ref{#1})}
\newcommand{\sref}[1]{\S\ref{#1}}

\newcommand{\fref}[1]{Figure \ref{#1}}
\newcommand{\btab}[1]{\begin{tabular}{#1}}
\newcommand{\etab}{\end{tabular}}

\newcommand{\comment}[1]{}

\newcommand{\qed}{\nobreak \ifvmode \relax \else
      \ifdim\lastskip<1.5em \hskip-\lastskip
      \hskip1.5em plus0em minus0.5em \fi \nobreak
      \vrule height0.75em width0.5em depth0.25em\fi}